**Topical Review: optics of exciton-plasmon nanomaterials**

Maxim Sukharev[1] and Abraham Nitzan[2]

[1]College of Integrative Sciences and Arts, Arizona State University, Mesa, Arizona 85212, USA

[2]Department of Chemistry, University of Pennsylvania, Philadelphia, Pennsylvania 19104, USA

**Abstract**

This review provides a brief introduction to the physics of coupled exciton-plasmon systems, the theoretical description and experimental manifestation of such phenomena, followed by an account of the state-of-the-art methodology for the numerical simulations of such phenomena and supplemented by a number of FORTRAN codes, by which the interested reader can introduce himself/herself to the practice of such simulations. Applications to CW light scattering as well as transient response and relaxation are described. Particular attention is given to so-called strong coupling limit, where the hybrid exciton-plasmon nature of the system response is strongly expressed. While traditional descriptions of such phenomena usually rely on analysis of the electromagnetic response of inhomogeneous dielectric environments that individually support plasmon and exciton excitations, here we explore also the consequences of a more detailed description of the molecular environment in terms of its quantum density matrix (applied in a mean field approximation level). Such a description makes it possible to account for characteristics that cannot be described by the dielectric response model: the effects of dephasing on the molecular response on one hand, and nonlinear response on the other. It also highlights the still missing important ingredients in the numerical approach, in particular its limitation to a classical description of the radiation field and its reliance on a mean field description of the many-body molecular system. We end our review with an outlook to the near future, where these limitations will be addressed and new novel applications of the numerical approach will be pursued.

## 1. Introduction

The collective response of systems involving interacting material systems and optical fields has been studied for a long time, yielding many fascinating modes of behavior. Traditionally light is regarded only as a probing and triggering tool and the observed response is associated with properties of the material system. It has however long been known that often an observed signal reflects not properties of the material system but of the light-matter hybrid.[i] The most explicit manifestations of this hybrid nature are polaritons – hybrid states of light and matter that display properties of both. Their classical description is obtained when solving the Maxwell equations in an environment whose dielectric properties, represented by the dielectric response function, $\varepsilon(\mathbf{k}, \omega)$ reflect resonance oscillations of charge or dipole fields. This review follows the common practice of using the position-local dielectric response (that is, replacing $\varepsilon(\mathbf{k}, \omega)$ by $\varepsilon(\omega) = \varepsilon(k = 0, \omega)$), which is experimentally accessible for many systems, however it should be stated at the outset that non-local effects may be important and understanding the extent of their importance continues to be a challenge. With regard to the local dielectric function, one can use the numerical data obtained experimentally, or commonly used model dielectric functions with parameters chosen to approximate such the numerical data.

*Dielectric response models.*

For metals, the simplest description of dielectric response is provided by the Drude model[1, 2] that considers conducting electrons as an ideal gas of non-interacting charged particles moving under Newton's second law augmented by including a phenomenological friction term (see, e.g. Refs. [3, 4]). This leads to the dielectric response function in the form

$$\varepsilon_r(\omega) = \varepsilon_{r,\infty} - \frac{\omega_p}{\omega(\omega - i\gamma)}. \tag{1}$$

Here $\omega_p$ is the plasma frequency, $\omega_p = n_e e^2 / (\varepsilon_0 m_e) \sim 10^{16} \, \mathrm{s}^{-1}$, where $n_e$ and $m_e$ are the electron mass and the electrons density, respectively, and $\varepsilon_{r,\infty}$ represents high frequency contributions from atomic core electrons (we use the notation $\varepsilon_r = \varepsilon / \varepsilon_0$ throughout this review). Applying this model to light propagation in homogeneous bulk metals, the dispersion relation $k^2 = \omega^2 \varepsilon_r(\omega) / c^2$ becomes for $\omega \gg \gamma$ (which holds near and above $\omega_p$) $k = (\omega/c)\sqrt{\varepsilon_{r,\infty} - \omega_p^2 / \omega^2}$, implying that electromagnetic wave propagation is possible only for $\omega\sqrt{\varepsilon_{r,\infty}} > \omega_p$, while at $\omega\sqrt{\varepsilon_{r,\infty}} = \omega_p$ all conducting electrons throughout the metal oscillate in phase. Note that this dispersion appears (see Fig. 1) to result from avoided crossing between the metal plasmon mode of frequency $\omega_p$ and the light line $k = \omega/c$ that represent this dispersion well for $\omega \gg \omega_p$. The distortion of this dispersion as $\omega \to \omega_p$ from above indicates strong coupling between light and plasma charge oscillations in the metal, creating the hybrid plasmon (or plasmon-polariton) mode.

---

[i] It should be emphasized that this distinction is sometimes in our mind only. Spontaneous emission of an excited molecule, which is often regarded as a molecular property, is a property of the molecule-field system in the same way that radiationless relaxation of this molecule is a property of the molecule and its environment.

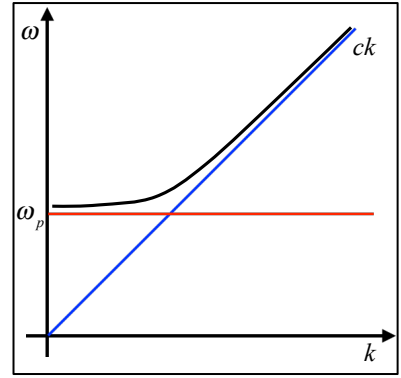

**Figure 1**. The dispersion $\omega(k)$ (black line) for electromagnetic propagation in a Drude metal appears to result from avoided crossing between the plasmon frequency $\omega_p$ (red line) and the light line $\omega = ck$ (blue line).

In general the Drude model qualitatively describes the optical response of metals for a short range of frequencies. When modeling scattering of electromagnetic radiation by metal systems it is a common practice to treat $\omega_p$, $\varepsilon_{r,\infty}$, and $\gamma$ as fitting parameters to better describe real and imaginary parts of the dielectric function for a given range of frequencies. Following this approach, another commonly used model combines the Drude function with the more general Lorentz function[5]

$$\varepsilon_r\left(\omega\right) = \varepsilon_{r,\infty} - \frac{\omega_p^2}{\omega^2 - i\gamma\,\omega} - \sum_{n=1}^{N}\frac{\Delta\varepsilon_{r,n}\omega_n^2}{\omega^2 - i\gamma_n\omega - \omega_n^2}, \qquad (2)$$

where the third term is the sum over contributions referred to as arising from the $n^{\text{th}}$ Lorentz oscillator. It can represent the effect of interband transitions as well as motions of the ionic cores. Parameters in the Lorentz part are treated as phenomenological fitting parameters. For example, it was shown that the optical response of silver could be described by (2) for the range of frequencies between 0.1 eV and 5 eV with 5 Lorentz poles.[5]

A particular form of the Lorentz function is often used to describe the near infrared response of ionic crystals, where a simple model for the dielectric response yields[3]

$$\varepsilon_r\left(\omega\right) = \varepsilon_{r,\infty} + \frac{\varepsilon_{r,\infty} - \varepsilon_{r,0}}{\left(\omega/\omega_T\right)^2 - 1} = \frac{\varepsilon_{r,\infty}\left(\omega/\omega_T\right)^2 - \varepsilon_{r,0}}{\left(\omega/\omega_T\right)^2 - 1} = \varepsilon_{r,0}\frac{\left(\omega/\omega_L\right)^2 - 1}{\left(\omega/\omega_T\right)^2 - 1} \qquad (3)$$

where $\varepsilon_{r,0}$ and $\varepsilon_{r,\infty}$ are, respectively, the static $\left(\omega \to 0\right)$ and the electronic $\left(\omega_T \ll \omega < \omega_{el}\right)$ dielectric response constants, while $\omega_T$ and $\omega_L = \omega_T\sqrt{\varepsilon_0/\varepsilon_\infty}$ are the long wavelength limits of the transversal and longitudinal optical phonon frequencies. Using again $k = \left(\omega/c\right)\sqrt{\varepsilon_r\left(\omega\right)}$ we see that no light can propagate in the crystal for $\omega_T < \omega < \omega_L$ because ε is negative in this region. (note that $\omega_L > \omega_T$ follows from the fact that the full dielectric response $\varepsilon_{r,0}$ is larger than $\varepsilon_\infty$ which reflects electronic response only). For $\omega \to 0$ $k \to \left(\omega/c\right)\sqrt{\varepsilon_{r,0}}$ while in the limit $\omega \to \infty$ $k \to \left(\omega/c\right)\sqrt{\varepsilon_{r,\infty}}$.

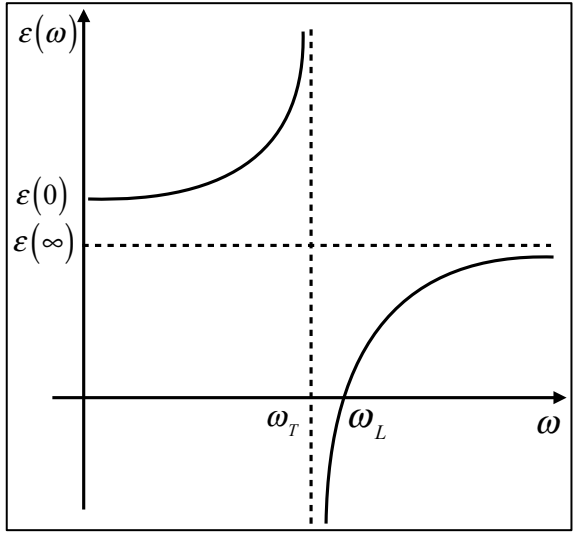

The overall dispersion as shown schematically in Fig. 2 can be interpreted as showing an avoided crossing between the low frequency light line and the transversal optical phonon as $\omega \to \omega_T$ from below, and a crossing between the high frequency light line and the longitudinal optical phonon as $\omega \to \omega_L$ from above. The distortion of the light line near the optical phonon frequencies and the formation of a frequency gap are again manifestation of strong coupling between the radiation field and the lattice optical phonons that results in hybrid phonon-polariton modes.

**Figure 2**. Dielectric constant as a function of frequency for a diatomic ionic crystal.

A qualitatively similar behavior is encountered at higher frequencies when the frequency dependence of the dielectric response is dominated by electronic resonances. A simple model for the dielectric function of a system of randomly oriented molecules is[ii]

$$\varepsilon_r\left(\omega\right) = \frac{\omega - \omega_L}{\omega - \omega_T}. \qquad (4)$$

Here $\omega_T = \omega_0 - \eta$ where $\eta = \left|\mu\right|^2/3\varepsilon_0\hbar\upsilon$ is of order $10^{13}$ s$^{-1}$ in condensed systems and $\omega_L = \sqrt{\varepsilon_{r,0}}\,\omega_T$ (here $\varepsilon_{r,\infty} = 1$ and $\varepsilon_{r,0} > 1$) are again frequencies of transversal and longitudinal exciton modes, respectively. As in Eq. (3), $\varepsilon$ diverges at $\omega = \omega_T$, vanish at $\omega = \omega_L$ and shows a forbidden gap for $\omega_T < \omega < \omega_L$ where EM waves cannot propagate because $\varepsilon < 0$. The dispersion relation for the resulting exciton-polariton is similar to that shown in Fig. 2, except that $\omega_T$ and $\omega_L$ now take values of the order of electronic transition frequencies $\sim 10^{15}$ s$^{-1}$.

*Collective optical response of metallic nanostructures.*

---

[ii] Eq. (4) is obtained from the polarizability expression $\alpha_{el}\left(\omega\right) \sim \left|\mu\right|^2/\hbar\left(\omega_0 - \omega\right)$ where $\mu$ is the transition dipole matrix element between the two electronic states and $\omega_0$ is the corresponding transition frequency (energy spacing between these states), and the Clausius-Mossotti equation, $\dfrac{\varepsilon - 1}{\varepsilon + 2} = \dfrac{\alpha}{3\upsilon}$ where $\upsilon$ is the volume per molecule (inverse molecular number density)

Geometry strongly affects the way by which such collective response phenomena are manifested. At a planar surface between two dielectrics characterized by dielectric response functions $\varepsilon_1(\omega)$, $\varepsilon_2(\omega)$ (see Fig. 3) Maxwell's equation can yield a solution that propagate along the surface, $\exp\left(ik_x x - k_{zj}|z| - i\omega t\right); j = 1,2$ with

$$k_x = \frac{\omega}{c}\sqrt{\left(\frac{\varepsilon_1\varepsilon_2}{\varepsilon_1+\varepsilon_2}\right)} \quad \text{and} \quad k_{zj} = \frac{\omega}{c}\sqrt{\frac{\varepsilon_j^{\,2}}{\varepsilon_1+\varepsilon_2}} \qquad (5)$$

offering a variety of behaviors for different interfaces, namely different choices of $\varepsilon_j(\omega)$, $j = 1, 2$.[6, 7] For a recent review see Ref. [8]. Importantly, when the two sides of the interface have their own collective response character, e.g. metal plasmons and molecular excitons, the resulting surface modes can display interaction between such collective excitations, offering new modes of behaviors and potentially novel applications.[8]

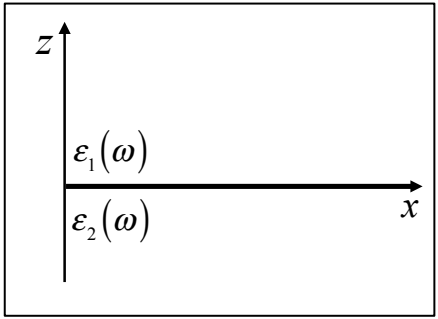

**Figure 3**. Planar interface between two dispersive dielectrics.

Such surface modes, including excitations obtained on more structured surfaces such as periodic arrays of holes or slits[9] are referred to as surface polaritons, and in the common case involving metal surfaces - surface plasmon polaritons (SPPs). Small particles constitute another class of systems with interesting plasmon behavior. For a small metal particle, plasmonic charge oscillations are obviously not a traveling wave but a collective electronic response localized on the particle's surface. Such excitations are referred to as localized surface plasmon-polaritons (LSPPs).[10] For a comprehensive review of terminology used in plasmonics see Chapter 3 in [11]. While the standard description is found in many textbooks, we use here an approximate representation[12, 13] that can serve to illustrate the connection to other electromagnetic response situations. Limiting ourselves to a small metal sphere in the electrostatic limit (radius $a$ and all distances considered much smaller than the radiation wavelength), the potential associated with the plasmon excitations is given by[iii]

$$\Phi(r,\theta,\phi) = \sum_{l=1}^{\infty}\sum_{m=-l}^{l}\left(A_{lm}(r,\theta,\phi)b_{lm} + A_{lm}^*(r,\theta,\phi)b_{lm}^*\right) \qquad (6)$$

where the functions A are defined in terms of the spherical harmonics $Y_{lm}$

---

[iii] In the electrostatic limit the interaction of a molecular dipole $\boldsymbol{\mu}$ is well represented by $-\boldsymbol{\mu}\cdot\boldsymbol{E} = \boldsymbol{\mu}\cdot\nabla\Phi$

$$A_{lm}(r,\theta,\phi) = \left[\frac{\hbar}{al\,\mathrm{Re}\,\varepsilon'(\omega_l)}\right]^{1/2} Y_{lm}(\theta,\phi)\begin{cases}\left(\dfrac{r}{a}\right)^{-l-1} & r \geq a \\[2mm] \left(\dfrac{r}{a}\right)^{l} & r \leq a\end{cases} \tag{7}$$

with $\varepsilon'(\omega) = d\varepsilon/d\omega$. In this representation, the free plasmon modes $b_{lm}$ satisfy the equations of motion of damped harmonic oscillators, $\dot{b}_{lm} = -i\omega_l b_{lm} - (1/2)\gamma_l b_{lm}$, where $\omega_l$ is the solution of

$$\mathrm{Re}\,\varepsilon_r(\omega_l) = -\frac{l+1}{l}, \tag{8}$$

while $\gamma_l$ is given by

$$\gamma_l = \frac{2\,\mathrm{Im}\,\varepsilon(\omega_l)}{\mathrm{Re}\,\varepsilon'(\omega_l)}. \tag{9}$$

The derivation of this form, and in particular assigning harmonic nature to the particle plasmons assume that these plasmons are underdamped and well separated, which holds only for the low $l$ plasmons of some metals. When these conditions hold, we can if needed consider the quantization of this field by replacing $b$ and $b^*$ by raising and lowering operators $\hat{b}$ and $\hat{b}^\dagger$. It should be emphasized however that while the representation of plasmons in small metal particles as (damped) harmonic oscillations is a common practice, this is more a matter of convenience than a rigorous result. In fact, experimental evidence that plasmon excitations in metal particles can be saturated,[14] implies that their assigned harmonic nature is only an approximation.

Another important geometry is that of a cavity, where the radiation field is confined within reflecting walls. When such a cavity supports one mode of frequency ω the field at position $x$ along the cavity axis is given by

$$\mathcal{E}(t) = i\sum_{\mathbf{k}}\sum_{\boldsymbol{\sigma}_{\mathbf{k}}}\sqrt{\frac{\hbar\omega_k}{2\varepsilon\Omega}}\boldsymbol{\sigma}_{\mathbf{k}}\left(a_{\mathbf{k},\boldsymbol{\sigma}_{\mathbf{k}}}e^{i(\mathbf{k}\cdot\mathbf{r}-\omega_k t)} - a_{\mathbf{k},\boldsymbol{\sigma}_{\mathbf{k}}}^*e^{-i(\mathbf{k}\cdot\mathbf{r}-\omega_k t)}\right) \tag{10}$$

with the classical amplitude $a$ ($a^*$) becoming the raising and lowering operators $\hat{a}$ and $\hat{a}^\dagger$ in the quantum theory. Importantly, interactions of such mode with other degrees of freedom depend on the cavity volume $\Omega$, namely the cavity length $L_x$ ($L_y$ and $L_z$ are assumed to be much smaller than the wavelength). In the ideal case of perfect mirrors, the metal is not involved except by imposing the reflecting boundary conditions. Optical interactions in high-Q cavities can be analyzed in terms of such cavity modes, adding when needed small damping into the time dependence of Eq. (10). In optical regimes where the loss is higher the objects making the mirror become part of the system, cavity physics becomes inter-metal gap physics, and the modes responsible for optical excitations in the space between metals become plasmon-polariton modes whose properties are derived from the dielectric properties of the metals and the geometries of both the gaps and the metal objects. To understand the electromagnetic field behavior in such gaps one needs to consider the full collective response of the system comprising the metal object and the space between them. For simple geometries this leads to considerations of plasmon hybridization,[15-18] (see Fig. 4 below) while in more complex systems numerical solutions of the Maxwell equations are needed.[19]

*The molecular system.*

Next, consider the molecular system. Many observed phenomena could be analyzed, at least qualitatively, modeling the molecule as a polarizable point object. In this physical limit the effect of the molecule on the plasmonic system is largely disregarded - it is taken into account only so much as it affects the molecule itself (for example by creating image effects or, in more sophisticated treatments, by adding self energy terms into the molecule equations of motion). In this limit the main effect of proximity to a metal surface is assumed to be the molecular interaction with the local electromagnetic field[iv] that reflects the response of the metal system to the electromagnetic field of the molecule as well to any incident radiation.

Many interesting observations can be understood qualitatively and at time quantitatively on this level. This includes the electromagnetic theory of surface enhanced Raman scattering;[26] for recent reviews see Refs. [27, 28] that has been later generalized to other photophysical and photochemical properties of molecules adsorbed on metal surfaces and nanostructures,[12, 29-31] and analysis of observed modification of radiative and non-radiative relaxation rates of molecules seated near such structures. For example, for a molecule represented by an oscillating charge distribution near a small dielectric structure, the modification of the radiative relaxation rate can be obtained[12, 13, 32] by evaluating the total oscillating dipole in the metal-molecule composite (for a general analysis of such observations near a flat metal surface see Ref. [33]) . For a molecule represented by an oscillating point dipole at a distance R from the center of a small dielectric sphere of radius $a$, where all distances correspond to the electrostatic limit, this leads to the radiative relaxation rate

---

[iv] At close proximity, other effects such as electron transfer between metal and molecules that affect not only the electromagnetic response but also the electronic structure of the metal-molecule composite come into play. [20], [21], [22], [23], [24], [25] We do not deal with such effects in this review.

$$\Gamma_R = |B|^2 \Gamma_{R0},$$ (11)

where $\Gamma_{R0}$ is the corresponding rate for the free molecule and where

$$B = \begin{cases} 1 + \dfrac{2(\varepsilon - \varepsilon_s)a^3}{(\varepsilon + 2\varepsilon_s)R^3} & \perp \\[4ex] 1 - \dfrac{(\varepsilon - \varepsilon_s)a^3}{(\varepsilon + 2\varepsilon_s)R^3} & \| \end{cases}$$ (12)

In Eq. (12) $\varepsilon = \varepsilon(\omega_m)$ and $\varepsilon_s = \varepsilon_s(\omega_m)$ are the dielectric response functions of the sphere and its environment, respectively, evaluated at the molecular transition frequency $\omega_m$, and (here and below) $\perp$ and $\|$ denote configurations with the dipole is perpendicular and parallel to the sphere surface. A possibly large effect is seen when $\omega_m$ is close to $\omega_p$, the dipolar plasmon frequency that satisfies $\text{Re}\left[\varepsilon(\omega_p) + 2\varepsilon_s(\omega_p)\right] = 0$, provided that the imaginary part of this sum is not too large.[v] The latter situation that is known to exist for the coinage metals (Ag, Au, Cu) as well as Al. The non-radiative relaxation rate can be calculated by evaluating the rate at which the oscillating molecular charge distribution drives heat generation in the metal, $\int d^3r (1/2)\sigma|E|^2$ where $\sigma = \omega\,\text{Im}\left(\varepsilon(\omega_m)\right)$ is the conduction at the driving frequency. This leads to

$$\Gamma_{NR}^{\perp} = -\frac{|\mu|^2}{8\pi\varepsilon_0\hbar a^3} \sum_{l=1}^{\infty} (l+1)^2 (2l+1) \left(\frac{a}{R}\right)^{2(l+2)} \text{Im}\frac{1}{l\varepsilon_r + (l+1)\varepsilon_{rs}},$$ (13)a

$$\Gamma_{NR}^{\|} = -\frac{|\mu|^2}{16\pi\varepsilon_0\hbar a^3} \sum_{l=1}^{\infty} l(l+1)(2l+1) \left(\frac{a}{R}\right)^{2(l+2)} \text{Im}\frac{1}{l\varepsilon_r + (l+1)\varepsilon_{rs}}.$$ (13)b

The sums in these expressions represent the contributions of different multipole plasmons to the non-radiative relaxation rate, and are seen to be enhanced at the corresponding plasmon resonance frequencies for which $\text{Re}\left[l\varepsilon(\omega) + (l+1)\varepsilon_s(\omega)\right] = 0$. Note that these rates represent the non-radiative relaxation induced by the metal, and will add to the rates of other non-radiative processes that may be present in the molecular system. In fact, at close proximity, another non-radiative decay channel induced by the metal, electron tunneling between metal and molecule, should also be taken into account.

---

[v] In standard applications $\varepsilon_s$ is considered real

Validity limits of the model approximations made above should be kept in mind when considering applications to real molecular/metal systems.[34-44] (See [45],[46] for recent reviews) A point dipole can hardly represent a large dye molecules adsorbed on a small metal particle. The use of a local dielectric function to represent the dielectric response of small particles become questionable for small particle sizes (see Refs [47-50]). At large particle size and/or molecule-distances the electrostatic approximation breaks down and a full retarded electromagnetic theory is needed. This fact, overlooked in some interpretations of distance dependence of molecule-metal particle energy transfer,[41, 42, 45, 46, 51-54] was highlighted in a recent numerical calculation.[55] Many other generalizations of the theory have been published in the past three decades.[36, 48, 56-73]

It is notable that the Gersten-Nitzan theory as many of its generalizations represent the molecule as a classical dipole.[vi] This may appear strange considering the fact that the rate of spontaneous emission, an inherently quantum phenomenon, is obtained from this calculation. The well-known[12, 13, 74-79] reason is that this calculation addresses only the effect of proximity to a metal surface, namely the *ratio* between the molecular radiative lifetimes at and away from the metal surface, and this ratio does not depend in our simple model on the quantum nature of this process. Another way to understand this is to observe that the radiative relaxation rate can be obtained from the golden rule formula in terms of the local density of final states of the radiation field, a property that can be calculated for different positions in an inhomogeneous dielectric environment using information obtained from the classical Maxell equations.[80, 81]

Finally, and closer to the focus of this review, in most observations of interacting molecules-plasmon system, the molecular system is not a single molecule but an assembly of atoms or molecules – a molecular aggregate or a molecular layer adsorbed on the metal surface. This has two important consequences: First, the molecular subsystem can have a substantial effect on the plasmonic response of the metal, so the system optical behavior has to be considered by treating its molecular and metal parts self-consistently. Second, coherent response of the molecular assembly can play an important role in the ensuing dynamics, and question about what determines coherence and dephasing and their impact on the molecular optical response naturally comes up. For a large molecular assembly, for example when one of the dielectrics separated by a planar interface is the molecular environment, the coherent response of the molecular system can be accounted for by using a suitable dielectric function to describe this environment as in Eq. (5). Alternatively, a microscopic description of the molecular side has the advantage that it can describe processes that go beyond simple (linear) dielectric response such as dephasing and nonlinear spectroscopy. For this level of description the molecule can be modeled as an n-level system (most simply

---

[vi] this is true also for many treatments of SERS, with the additional attribute that the dipole moment is assume to depend on the nuclear coordinate(s)

n = 2, see, e.g. Refs. [82-85] and be described by the optical Bloch equations (Chapter 10 in Ref. [86]) driven by the local electromagnetic field. The latter, in turn, is evaluated from the classical Maxwell equations in which the time derivative of expectation value of the molecular dipole, $\text{Tr}\left(\hat{\mu}\hat{\rho}\right)$ constitutes a current source. This level of calculations is important in particular when many molecules are involved and we return to its detailed description below.

*Collective response of molecule-metal nanostructures.*

An important attribute of the optical response of molecular aggregates is their collective optical response. Consider first an atomic or molecular assembly without the metal component. Systems relevant to the present discussion are encountered in optical studies of molecular dye aggregates[87] on one hand, and in studies of optical properties of low dimensional (layers and dots) semiconductor structures[88, 89] on the other, although other interesting system such as dye containing nicrodroplets can be investigated along similar lines.[90-93] The simplest model pertaining to our applications are studies of the optical response of assemblies of 2-level atoms.[94-99] In a classic paper, Dicke [100] has shown that, because of their mutual interaction with the radiation field, a cluster of emitters has some modes that can emit cooperatively with rates that scale with the number of atoms, so called superradiant modes, while other modes are only weakly coupled to the radiation field. Closely related is superfluorescence, [101] where a system of initially uncorrelated excited emitters develops coherence through their mutual interaction with the radiation field. Prasad and Glauber [94-97] have analyzed in detail the emission properties of such slabs and spheres of two-level emitters under conditions of (a) weak excitation (namely, the system remains close to its ground state so only its linear response is considered; (b) continuum approximation - assuming the system is dense enough to allow smoothening of its atomistic structure and consider it as a continuous polarizable medium; (c) timescale separation between the atomic resonance frequency (the spacing between the atomic levels) and the emission dynamics, as well as between this dynamics and the time it take light to cross the system; and (d) phase destroying processes such as Doppler or collisional broadening are ignored. Under these assumptions they were able to describe the emission properties of such systems in terms of eigenstates of an integral equation for the system polarization, showing that a few of these eigenstates behave as superradiant modes, emitting at a rate that scales linearly with the number of atoms. All other modes have zero (in the limit of small clusters) decay rates or infinite lifetimes. Similarly, Svidzinsky and coworkers [98] have studied an N-atom cluster of linear size $R$ with only one atom excited in the initial state, and have shown that in the small cluster limit ( $k_0 R \to 0; k_0 = \omega_0 / c$ where $\omega_0$ and $c$ are the resonance frequency and the speed of light, respectively) only one state of this system can emit with rate $\sim \gamma_R N$ ($\gamma_R$ being the radiative decay rate of a single excited atom) while the others have infinite lifetimes. In the opposite case of large clusters there are many superradiant states with relaxation rates $\sim \gamma_R N / \left( k_0 R \right)^2$. We note in passing that other types of coherent dynamics such as soliton-like exciton propagation,[102] lasing[103] and Bose-Einstein condensation[104] have been discussed. On the other hand, dephasing phenomena and disorder can change the qualitative behavior of such systems, and energy transfer in excitonic systems can become diffusive on timescales shorter than their radiative emission, with typical coherence lengths of the order of $\sim 100$ nm. (See e.g. [105, 106]).

Next, consider the interaction of such an aggregate of emitters with a plasmon-sustaining interface. (Ref. [107] was perhaps the first publication to focus attention on such systems. See Ref. [108] for a recent review). In accord with the single emitter case discussed above, standard spectroscopic properties – absorption emission and lifetimes are modified.[109-121] Of particular interest are observations of plasmon effects on interaction between emitters. The simplest manifestation of such effects is the plasmon-induced enhancement of energy transfer between emitters.[62, 79, 122-148] This can be used in conjunction with metal configuration that supports sub-wavelength waveguiding to direct exciton propagation in predetermined ways.[149-163] This plasmon enhanced energy transfer implies that intersite excitonic coupling is effectively enhanced by coupling to the underlying plasmons, which implies that exciton-plasmon coupling should also affect superradiance and superfluorescence from excited emitter aggregates. Refs. [70, 71, 134, 164-171] have shown that interatomic (or intermolecular) interaction via coupling to plasmon can dominate the formation of cooperative response over the standard coupling to the free radiation field, and leads to qualitatively similar behavior as described above –a few modes of the molecular/atomic aggregate are superradiant while the others can be strongly sub-radiant. Plasmonic metal structures were indeed shown to enhance the collective nature of emission from emitter aggregates.[172]

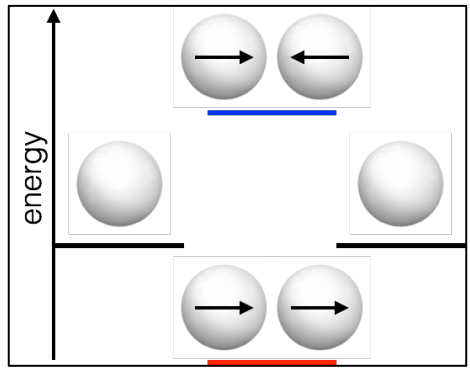

**Figure 4**. The plasmon hybridization model applied to two interacting nanoparticles. The model explains the physical nature of two observed resonances corresponding to anti-bonding (upper state) and bonding (lower state) formed when two nanoparticles are placed close together.

While the above discussion has distinguished between inter-emitter coupling induced by mutual coupling to the radiation field or to surface plasmons, it is important to emphasize that the physics of these phenomena is essentially the same, and the dominance of coupling to plasmon mode results from the focused local nature of the latter. All phenomena associated with exciton-plasmon coupling, including enhanced cooperative emission and enhanced exciton transport, have their analogs in excited molecules and molecular assemblies interacting with optical modes localized in cavities and other nanostructures such as photonic crystals, where molecule-radiation field coupling can be tuned by geometrical characteristics of the systems.[173-175]

*Strong coupling*

The consequences of interaction between molecules and plasmon-supporting metal surfaces have so far been discussed in terms of modified molecular properties – absorption, scattering, and relaxation rates. A growing attention is given in recent years to another manifestation of this interaction (and the equivalent phenomenon in systems comprising molecules or semiconductor quantum dots and cavity-confined radiation field modes) loosely referred to a "strong coupling", where the focus is on the hybridization between the material and optical modes to form hybrid states that comprise both.[176] An example is shown in Fig. 4. Two localized surface plasmon-polariton (LSPP) modes of two nanoparticles are formed by combining the LSPP modes of a spherical nanoparticle corresponding to bonding and anti-bonding states. Treating plasmons as incompressible deformations of electron gas and calculating the total electrostatic energy of the system one obtains a set of coupled equations on amplitudes of deformation fields describing LSPP modes of the combined system.[15, 18]

An often encountered model used to explain this aspect of intersystem coupling comprises two coupled classical harmonic oscillators *A* and *B*.[177] The corresponding Hamiltonian

$$H = \frac{P_A^{\ 2}}{2m_A} + \frac{P_B^{\ 2}}{2m_B} + \frac{1}{2}k_A X_A^{\ 2} + \frac{1}{2}k_B X_B^{\ 2} + \frac{1}{2}\kappa\left(x_A - x_B\right)^2 \tag{14}$$

can be expressed in terms of two independent normal modes of frequencies

$$\omega_\pm^{\ 2} = \frac{1}{2}\left(\omega_A^{\ 2} + \omega_B^{\ 2} \pm \sqrt{\left(\omega_A^{\ 2} - \omega_B^{\ 2}\right)^2 + \frac{4\kappa^2}{m_A m_B}}\right) \tag{15}$$

with

$$\omega_N = \sqrt{\frac{k_N + \kappa}{m_N}}; \ \ N = A, B \tag{16}$$

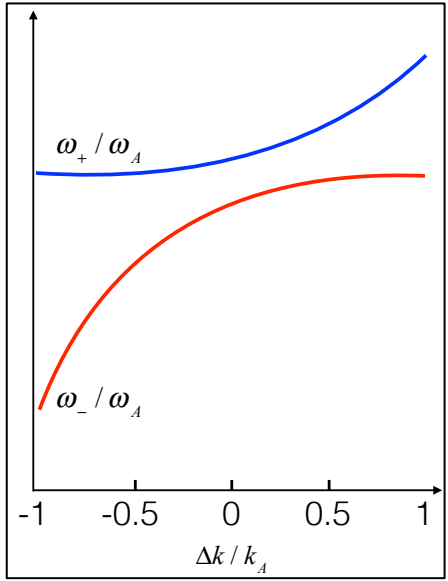

**Figure 5**. The normal mode frequencies of a system of two linearly coupled harmonic oscillators, plotted against the difference $\Delta k = k_B - k_A$ between the force constants of the free oscillators

Plotting the frequencies $\omega_+$ and $\omega_-$ against $\omega_B$ while keeping $\omega_A$ constant shows a typical coupling induced avoided crossing: When $\kappa = 0$ the two lines are: a horizontal one, $\omega_+ = \omega_A(\kappa = 0)$ indicating that this root does not depend on $\omega_B$ and the line $\omega_- = \omega_B(\kappa = 0)$ with slope 1 that correspond to the linear dependence of this root on $\omega_B$ obviously cross at $\omega_B = \omega_A$. When $\kappa \neq 0$ the lines avoid each other and achieve minimum difference $\left(\omega_+{}^2 - \omega_-{}^2\right)_{\min} = 2\kappa/\sqrt{\sqrt{m_A m_B}}$ and strong mixing when $\omega_B = \omega_A$. A familiar analogous situation is the two-level emitter in quantum mechanics, where diagonalizing the Hamiltonian

$$
\begin{aligned}
\hat{H} &= \hat{H}_0 + \hat{V}, \\
\hat{H}_0 &= E_A \left| A \right\rangle \left\langle A \right| + E_B \left| B \right\rangle \left\langle B \right|, \\
\hat{V} &= V_{AB} \left| A \right\rangle \left\langle B \right| + V_{BA} \left| B \right\rangle \left\langle A \right|
\end{aligned}
\tag{17}
$$

yield the eigenvalues

$$
E_{\pm} = \frac{E_A + E_B \pm \sqrt{(E_A - E_B)^2 + 4\left| V_{AB} \right|^2}}{2}
\tag{18}
$$

with a minimum spacing, the so called Rabi splitting, $2\left| V_{AB} \right|$. Yet another familiar system is the Jaynes-Cumming model [178] that describes coupled two-level system and a quantum harmonic oscillator

$$
\hat{H} = \hbar \omega \hat{a}^{\dagger} \hat{a} + E_1 \left| 1 \right\rangle \left\langle 1 \right| + E_2 \left| 2 \right\rangle \left\langle 2 \right| + \frac{1}{2} U \left( \hat{a}^{\dagger} \left| 1 \right\rangle \left\langle 2 \right| + \hat{a} \left| 2 \right\rangle \left\langle 1 \right| \right)
\tag{19}
$$

This model appears more relevant to our notion of a molecule (represented by a two-level system) coupled to a plasmon (described as a harmonic oscillator). However, the conservation of excitation

$$
\left\langle \hat{a}^{\dagger} \hat{a} \right\rangle + \left\langle \left| 2 \right\rangle \left\langle 2 \right| \right\rangle = \text{constant}
\tag{20}
$$

that follows from (19), implies that the states of this Hamiltonian correspond to an infinite set of coupled two-level systems that are distinguished by the number of harmonic oscillator quanta that are present when the actual 2-level system is in state 1. Each such a pair is thus described by the Hamiltonian (17) where $\left|A\right\rangle=\left|1,n\right\rangle$, $\left|B\right\rangle=\left|2,n-1\right\rangle$ while $V=U\sqrt{n}$. A new element is that the Rabi splitting is here proportional to $\sqrt{n}$, namely the amplitude of the harmonic subsystem. In addition, from the discussion above, following, e.g., Refs. [94-100], $\hat{V}$ is also expected to scale like $\sqrt{N}$ where $N$ is the number of molecules within a volume much smaller that the radiation wavelength that are coherently involved in the process.

These models serve to understand observations of mode coupling in optical response of composite nanosystems, ranging from plasmon hybridization[vii] in nanoparticle systems,[15-18, 179-187] to molecules, molecular assemblies, and semiconductor nanodots interacting with plasmons excited in metal nanostructures.[55, 158, 188-202] as well as cavity modes. [178, 203-210] For recent reviews see Refs. [108, 175] and [211].

An example for this behavior is shown in Figure 6. In later parts of this review we focus on the computational and theoretical analysis of these phenomena in coupled exciton-plasmon systems. Fig. 6 demonstrates the qualitatively similar behavior in coupled plasmon systems comprising metallic nanostructures. Specifically we focus on a simple system made of two closely spaced gold nanowires with characteristic lengths less than the incident wavelength schematically depicted in Figure 6a, and consider the linear optical response of this system to two different excitations, namely a conventional incident plane wave and a local dipole source. When the system is excited by the plane wave polarized along the axis connecting two

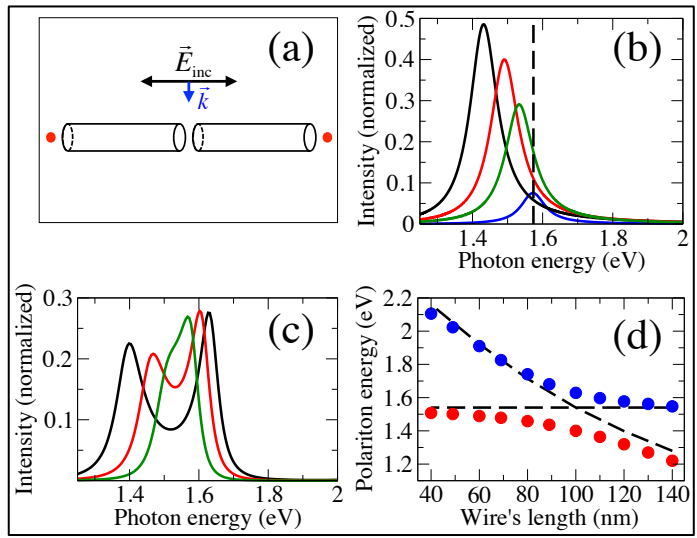

**Figure 6.** Two gold nanowires separated by a sub-diffraction gap. Panel (a) schematically depicts two wires. Two red dots indicate local dipole source and local detection point. This excitation scheme is compared to the conventional excitation by a plane wave. Panel (b) shows scattered EM intensity detected in the far field when the wires are excited by a plane wave. Blue line shows data for a single 100 nm long wire; black line shows data for the gap of 10 nm, red line – 20 nm, green line – 40 nm. Both wires are 100 nm long with a diameter of 20 nm. Panel (c) shows results when wires are excited by a local dipole placed 5 nm away from one of the wires. Different colors correspond to the same values of the gap as in panel (b). Panel (d) shows energies of hybrid plasmon states (red and blue dots) as functions of the length of one of the wires (another wire's length is fixed at 100 nm). Dashed curve corresponds to the energy of the longitudinal plasmon of a single wire.

---

[vii] The term 'hybridization' arises from the theoretical approach that qualitatively (and quantitatively in the linear regime) describes SPP modes supported by a complex system using a simple analogy with the hybridization of molecular orbitals in quantum chemistry.

wires the scattering intensity exhibits a single resonance that shifts to higher energies as the gap size increases (Fig. 6b). This is to be expected since the incident field excites longitudinal LSPP modes in each wire and their interaction strength drops as the gap widens. The energy of the resonant mode approaches a limit of a single wire as clearly seen in Fig. 6b. It is not surprising to see only a single mode since another hybrid state is the dark (anti-bonding) mode which is not accessible due to symmetry when the system is uniformly excited by the plane wave(see Fig. 4). Dark states however can be excited when a local dipole source is utilized (Fig. 6c). Here both bonding and anti-bonding hybrid states are seen in the transmission spectra. Using the language of the strong coupling physics the transmission spectra exhibit the Rabi splitting determined by the strength of the electromagnetic (EM) interaction between LSPP modes of each wire. Reflecting their relative energies (Fig. 4), the bonding and antibonding hybrid states are referred to as the lower and upper polaritons, respectively. The energy separation of these states increases as the gap between wires narrows, indicating the fact that the EM coupling strength increases. Fig. 6d explores the anticrossing behavior of the hybrid states when the length of one of the wires is varied. Without EM coupling between the wires the longitudinal LSSP mode of one wire would cross the energy of the LSSP mode of another as dashed lines in Fig. 6d illustrate. The coupling results in the avoided crossing behavior typical for strongly coupled systems.

We note that the observed Rabi splitting for the wires of 100 nm in length is 240 meV corresponding to the characteristic time of EM energy exchange between plasmonic particles of 17 fs. Such a fast dynamics calls for the utilization of plasmon modes (both bright and dark as we shall see below) as nano-optical probes[212] to investigate fundamentals of light-matter interaction.

Coming back to theoretical descriptions, to provide a complete qualitative understanding of these phenomena the models represented by Eqs. (14) and (17) have to be supplemented by proper descriptions of damping on one hand, and the probe used for the actual observation on the other. To show the consequence of taking these into account consider the model of Fig. 7 in which the two-level model of Eq. (17) is augmented by (a) having each of the two levels damped by coupling it to its own continuum and (b) having the absorption lineshape of this system probed by a photon coming from the far field and coupled to the system by some dipole operator $\mu$. As a primitive model to observation of such phenomena in coupled exciton-photon system level A can be a single exciton state and level B a single plasmon state (or a cavity mode), both accessible from the ground state (unexcited molecule-metal/cavity composite) and characterized by their mutual coupling V and lifetimes associated with their damping to their environments. (Typical plasmon lifetimes are 50 fs and those of cavity modes, determined by the cavity Q number, are much longer.) Exciton lifetimes are bounded from below by the radiative decay process, of order 1 ns in free space and possibly considerably longer in cavity environments, however such lifetimes are made considerably shorter by radiationless relaxation into environmental modes including the close-by metal interface). Denoted these damping rates by $\gamma_A$ and $\gamma_B$, the absorption lineshape is obtained in the form

$$
\begin{aligned}
L(\omega) &= \operatorname{Im} \frac{F}{\left(\hbar\omega - E_A + (i/2)\gamma_A\right)\left(\hbar\omega - E_B + (i/2)\gamma_B\right) - \left|V_{AB}\right|^2} \\
&= \operatorname{Im} \frac{F(\omega)}{\left(\hbar\omega - E_+\right)\left(\hbar\omega - E_-\right)} = \frac{F(\omega)\left(\hbar\omega - E_+^*\right)\left(\hbar\omega - E_-^*\right)}{\left|\left(\hbar\omega - E_+\right)\right|^2 \left|\left(\hbar\omega - E_-\right)\right|^2},
\end{aligned}
\tag{21}
$$

where

$$
\begin{aligned}
F(\omega) &= \left|\mu_{GA}\right|^2 \left(\hbar\omega - E_B + (i/2)\gamma_B\right) + \\
&\quad \left|\mu_{GB}\right|^2 \left(\hbar\omega - E_A + (i/2)\gamma_A\right) - \mu_{GA}V_{AB}\mu_{BG} - \mu_{GB}V_{BA}\mu_{AG}
\end{aligned}
\tag{22}
$$

and $E_\pm$ are given by Eq. (18) with $E_A$ and $E_B$ are replaced by $E_A - (i/2)\gamma_A$ and $E_B - (i/2)\gamma_B$, respectively.

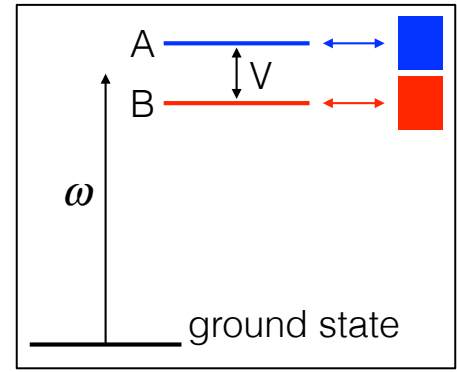

**Figure 7**. A model illustrating the consequence of coupling between two absorption resonances whose widths are determined by coupling to dissipating continua. At issue is the question whether their mutual coupling V is strong enough to keep them sufficiently apart so that two peaks are observed for all possible values of $E_A - E_B$.

When $E_A$ is far from $E_B$ ($\left|E_A - E_B\right| \gg \left|V_{AB}\right|$) the lineshape (21) clearly shows two maxima, near $\hbar\omega = E_A, E_B$. As $E_B$ changes and approaches $E_A$ these maxima come closer and may either coalesce into a single maximum or remain separated if the coupling $\left|V_{AB}\right|$ is large enough. At issue is the question whether this lineshape shows one or two maxima when they are their closest approach. For a quick estimate we can proceed with the following simplifications: take $\gamma_A = \gamma_B \equiv \gamma$ and assume that the structure of the lineshape is dominated by the denominator $D(\omega)$ of Eq. (21):

$$
D(\omega) = \left[\left(\hbar\omega - E_0 - \eta\right)^2 + (1/2)\gamma^2\right]\left[\left(\hbar\omega - E_0 + \eta\right)^2 + (1/2)\gamma^2\right],
\tag{23}
$$

where $E_0 = (1/2)\left(E_A + E_B\right)$ and $\eta = (1/2)\sqrt{\left(E_A - E_B\right)^2 + 4\left|V_{AB}\right|^2}$. The condition for this function to have two distinct minima is that it has a maximum at $\hbar\omega = E_0$, namely $\left[d^2D \middle/ d\omega^2\right]_{\omega = \hbar^{-1}E_0} < 0$. This translates into

$$\gamma < \sqrt{(E_A - E_B)^2 + 4|V_{AB}|^2}$$ (24)

so that the condition for a split peak at the closest approach $E_A = E_B$ is

$$\gamma < 2|V_{AB}|.$$ (25)

Situations where this condition is met are referred to as "strong coupling", however it should be kept in mind that "strong" implies that $|V_{AB}|$ is large relative to not only lifetime broadening as described above, but in fact to any other type of homogeneous and inhomogeneous broadening that can mask the Rabi splitting. To see the implications of this condition, consider first the coupling of a molecule to an unpopulated cavity mode. The coupling strength (Rabi frequency) implied by Eq. (10) is of order (taking $\varepsilon_r = 1$)

$$V = \mu\sqrt{\hbar\omega/2\varepsilon_0\Omega},$$ (26)

where $\mu$ and $\omega = 2\pi c / \lambda$ are the dipole matrix element and the frequency for the molecular transition and $\Omega$ is the cavity volume. Taking $\Omega = \lambda^3$ and using $\mu = 1$ Debye leads to Rabi frequency of the order of ~1 wavenumber ($10^{-4}$ eV). Compared to this, the damping rate of cavity modes can be relatively negligible, and radiative decay rates of molecular excitations are also 2-3 orders of magnitude smaller (and may be further reduced due to the cavity Purcell effect[213]). This implies that molecule-cavity mode systems can often be found in the strong coupling limit. When an aggregate of many molecules is involved the effective Rabi frequency may increase by 1-2 orders of magnitudes to the level observed in some experiments due to cooperative molecular response.

For a molecule interacting with the dipolar ($l = 1$) plasmon of a small metal nanoparticle we can make an equivalent consideration using Eq. (6) and $\mathbf{E} = -\nabla\Phi$ to obtain the molecule (again represented by its transition dipole $\mu$) – plasmon interaction. For a small molecule[viii] outside a metal sphere of radius $a$ at a distance $R > a$ from the sphere center perpendicular to the sphere surface this leads to[12, 13]

$$V = \mu\left(\frac{a}{R}\right)^3 \sqrt{\frac{3\hbar}{2\pi\varepsilon_0 \operatorname{Re}\varepsilon_r{}'\left(\omega_p\right)a^3}}$$ (27)

with $\varepsilon_r{}'(\omega) = d\varepsilon_r/d\omega$, $\omega_p$ is obtained from (cf. eq. (8)) $\operatorname{Re}\left(\varepsilon_r\left(\omega_p\right)\right) = -2$ and all distances are assumed small relative to $\lambda_p = 2\pi c / \omega_p$. Compared to Eq. (26) we see a similar structure with elements

---

[viii] The result (27) assumes a point-like molecule located at distance $R$ from the sphere center.

of differences: ω is replaced by $\varepsilon_r'\left(\omega_p\right)^{-1}$ and the cavity volume is replaced by the particle volume (a similar conclusion regarding the volume was reached in Ref. [201] following the analysis of Ref. [214]. In addition, the factor $\left(a/R\right)^3$ introduces an explicit dependence on the molecule-sphere distance.

Because $a$ can be far smaller than the radiation wavelength, the Rabi frequency near plasmonic particles can be considerable larger than that characterizing coupling to cavity modes. Indeed, for silver $\hbar\left(d\varepsilon/d\omega\right)_{\omega_p}^{-1}$ is of order 5 eV and for a molecule of $\mu = 10\,\text{D}$ near a silver sphere of radius $a = 4$ nm we find $V \sim 0.1\,\text{eV}$. On the other hand, the damping in this case is also much larger. It is usually dominated by the dipolar plasmon lifetime, of the order 50 fs for silver, corresponding to $\gamma \sim 0.1\,\text{eV}$.[ix] We conclude that strong coupling can be realized in such situations as well, in agreement with the observation of Ref. [201], although achieving it requires more fine-tuning of system parameters than in the molecule-cavity case.

Like the plasmon coupling example of Fig. 6, most demonstrations of strong coupling between molecules and optical modes have focused on the phenomenon of avoided crossing and similar manifestations (e.g. Fano lineshapes) in light scattering spectra. However other observations should be noted:

(a) Obviously, the energetic structure predicted by these models, as exemplified in Fig. 6, also implies the possibility to observe temporal (Rabi) oscillation in the time domain. Actual observations of such oscillations in ultrafast optical response of such systems have been rare. A recent work by Vasa, Lienau and co-workers describes such an observation.[198]

(b) The two level-harmonic oscillator models discussed above provide useful but by no means complete pictures of strongly coupled systems. In particular, the consequences of the vibrational motions in the molecular subsystems can be observed[215, 216] and should be included in a complete analysis of such strongly coupled systems.[217]

(c) In analyzing some of these examples of mode hybridization, the focus is on coupling between molecular excitations and optical/plasmon modes. At close proximity the possibility of electron tunneling between metal structures can considerably change the physics of the system and should be taken into account.[21, 22, 218] Similarly, electron transfer can strongly affect metal-molecule interaction at short distances and overshadows the effect of exciton-plasmon coupling. Indeed, plasmon-induced hot electron chemistry – processes in which following plasmon excitation electrons are exchanged between the metal

---

[ix] When the molecule sits directly on the metal surface energy and electron transfer to the metal provide other efficient relaxation routes with relaxation times possibly of similar order.

and the molecular system are currently drawing increasing attention.[219-235] (See also Refs [236-242] for recent theoretical work and Refs. [14, 243, 244] and [245] for reviews).

(d) Other important effects of such coupling have been observed and discussed. These include behaviors that reflect the modification of molecular electronic structure and dynamics,[246-252] spin relaxation,[253-255] metal work functions,[256] modification of electron and exciton transport properties because of interaction with cavity modes.[206, 207, 257] as well as nonlinear phenomena, e.g. induced transparency[258, 259] and second harmonic generation[260, 261] in coupled exciton-plasmon systems.

The simple theoretical considerations outlined above provide a qualitative understanding of many of the observed phenomena, however to understand the detailed behavior of actual experimental systems involving complex metal structures and an assembly of interacting molecules we need to resort to numerical studies. The rest of this review is aimed to provide a detailed account of the current state of the art of such studies.

## 2. Theoretical background for the numerical approach

Numerical simulations of atomic, molecular or semiconductor systems coupled to the radiation field have been done for a long time in different contexts.[101, 262-275] Including plasmon-sustaining metal nanostructures is a fairly recent development.[273] Below we refer to such systems as coupled exciton-plasmon systems. In what follows we describe the theoretical basis for simulating the dynamics and optical response of such systems, namely the ways by which the EM field, the molecular system, the metal, and the coupling between them are accounted for. The actual implementation of these descriptions in numerical solvers is described in the next section.

*The electromagnetic field*

In the vast majority of systems considered here the EM radiation can be treated classically using Faraday's and Ampere's laws

$$\frac{\partial \boldsymbol{B}}{\partial t} = -\nabla \times \boldsymbol{E},$$
$$\frac{\partial \boldsymbol{D}}{\partial t} = \frac{1}{\mu_0} \nabla \times \boldsymbol{B},$$

(28)

where $\mu_0$ is magnetic permeability of free space and the electric displacement field is defined as

$$\boldsymbol{D} = \varepsilon_0 \boldsymbol{E} + \boldsymbol{P}.$$

(29)

The macroscopic polarization, $\boldsymbol{P}$, is the essence of semiclassical theory of optical properties of exciton-plasmon materials. It is induced in a material subsystem by the EM field and in turn drives the EM field through the current density $\mathbf{J}(\mathbf{r},t) = \partial\mathbf{P}(\mathbf{r},t)/\partial t$.

*The metal*

In the simulations described below the metal is described as a continuum dielectric with a dielectric function given by the Drude (Eq. (1)) or Lorentz (Eq. (2) models with suitably chosen parameters. Such parameters for several metals are given in Ref. [5]. The advantage of using these models is that their effect can be described by adding auxiliary variables governed by Markovian dynamical equations to the description of the system dynamics. The specific equations that govern the dynamics of the current density depend on a choice of a dielectric function and are discussed in Section 3.

*The molecular system*

The simplest description of a molecular system accounts for its response to the radiation field in terms of a local dielectric function that can be described by a Lorentz model with suitably chosen parameters. With this approach, the numerical solution of coupled exciton-plasmon systems is reduced to solving the Maxwell equations in a continuous non-homogeneous dielectric. Such a description of the molecular system suffers from two shortcomings: First, it has to include some procedure to account for the molecular density and local orientation within the Lorentz parameters, and second, while it contains phenomenological damping parameters, it cannot distinguish between the effect of population relaxation (so called $t_1$ processes) and dephasing ($t_2$ processes), namely it cannot explicitly account for pure dephasing.

Both orientation effects and molecular density can be described in a classical microscopic description, in which individual molecules (or, in a coarse-grained description, different cells of a numerical grid) are described as Lorentz oscillators with suitably chosen parameters.[276] These oscillators, describing molecular dipoles, do not directly interact with each other. Rather, a mutual interaction is affected through their interaction with the EM field.[x]

The effect of pure dephasing, as a process different from excitation relaxation, can be accounted for by describing the molecular sub-system as a collection of quantum mechanical entities, albeit in a *mean-field approximation*. To describe this procedure we start with the Hamiltonian written as a sum of a molecular part $\hat{H}_M = \sum\limits_m \hat{h}_m$ , the radiation field $\hat{H}_R$ and their interaction $\hat{H}_{MR} = \sum\limits_m \hat{H}_{mR}$ , where the sums are taken over individual molecules.

---

[x] It is important to emphasize that this statement holds only for interaction within the frequency range accounted for by dynamical coupling with the classical electromagnetic field. Electrostatic interactions and van der Waals forces are also mediated by the ambient electromagnetic field but cannot be accounted for on the level of these simulations. If such interactions are needed they have to be added explicitly.

$$\hat{H} = \hat{H}_M + \hat{H}_R + \hat{H}_{MR}. \tag{30}$$

In $\hat{H}_{mR} = -\boldsymbol{E}\left(\boldsymbol{r}_m\right) \cdot \hat{\boldsymbol{\mu}}_m$ the electric field at the position of the molecule $m$ depends on the states of all other molecular dipoles. The molecular density matrix evolves according to the Liouville equation

$$\frac{d\hat{\rho}}{dt} = -\frac{i}{\hbar}\left[\hat{H}, \hat{\rho}\right]. \tag{31}$$

Now assume that $\hat{\rho} = \prod\limits_m \hat{\rho}_m$ with $\mathrm{Tr}_m \hat{\rho}_m = 1$ . Using this in (31) and taking trace over all molecules except $m$ leads to

$$\frac{d\rho_m}{dt} = -\frac{i}{\hbar}\left[\hat{h}_m - \left\langle \boldsymbol{E}\left(\boldsymbol{r}_m\right)\right\rangle \cdot \hat{\boldsymbol{\mu}}_m, \hat{\rho}_m\right] \tag{32}$$

with $\left\langle \boldsymbol{E}\left(\boldsymbol{r}_m\right)\right\rangle = \mathrm{Tr}'\left(\boldsymbol{E} \prod\limits_{m' \neq m} \hat{\rho}_{m'}\right)$ where $\mathrm{Tr}'$ represents trace over the states of all molecules except m.

The Liouville-von Neumann Equation (32) describes the dynamics of a single molecular dipole under the influence of the local average electric field at the position of this dipole. The latter includes the incident field and the average field of all other molecules. For a single molecule within an ensemble of two-level ( $\varepsilon_1, \varepsilon_2$ ) molecules, Eq. (32) yields (see also [277])

$$\frac{d\rho_{11}}{dt} = -i\frac{\boldsymbol{E}\boldsymbol{\mu}_{12}}{\hbar}\left(\rho_{12} - \rho_{12}^*\right),$$

$$\frac{d\rho_{12}}{dt} = i\left[\frac{\boldsymbol{E}\boldsymbol{\mu}_{12}}{\hbar}\left(\rho_{22} - \rho_{11}\right) + \Omega_{12}\rho_{12}\right], \tag{33}$$

$$\frac{d\rho_{22}}{dt} = i\frac{\boldsymbol{E}\boldsymbol{\mu}_{12}}{\hbar}\left(\rho_{12} - \rho_{12}^*\right),$$

where 1 and 2 denote the ground and the excited state, respectively, $\boldsymbol{\mu}_{12}$ is the corresponding transition dipole moment (assumed real), and $\Omega_{12} = \left(\varepsilon_2 - \varepsilon_1\right)/\hbar$ is the transition frequency. Another often encountered form of Eqs. (33) is obtained by defining $X = \rho_{12} + \rho_{12}^*$ , $Y = \rho_{12} - \rho_{12}^*$ and noting that $dX/dt = i\Omega_{12}Y$ . This leads to

$$\frac{d\rho_{11}}{dt} = -\frac{1}{\Omega_{12}}\frac{dX}{dt}\frac{\boldsymbol{E}\cdot\boldsymbol{\mu}_{12}}{\hbar},$$

$$\frac{d\rho_{22}}{dt} = \frac{1}{\Omega_{12}}\frac{dX}{dt}\frac{\boldsymbol{E}\cdot\boldsymbol{\mu}_{12}}{\hbar},$$ (34)

$$\frac{d^2X}{dt^2} = -\Omega_{12}^{\;2}X + 2\Omega_{12}\frac{\boldsymbol{E}\cdot\boldsymbol{\mu}_{12}}{\hbar}\left(\rho_{11}-\rho_{22}\right).$$

Next we (a) multiply these equations by the molecular density $n_M$ and the last equation also by $\boldsymbol{\mu}_{12}$, (b) denote the densities of molecules in states 1 and 2 by $n_M\rho_{11} = n_1;\;\; n_M\rho_{22} = n_2\;\;\left(n_1 + n_2 = n_M\right)$ and the polarization by $Xn_M\boldsymbol{\mu}_{12} = \mathbf{P}$, and (c) assume an isotropic system and average over all orientations of the molecular dipole to get $\left\langle\mu_x\mu_y\right\rangle = \left\langle\mu_x\mu_z\right\rangle = \left\langle\mu_y\mu_z\right\rangle = 0$ and $\left\langle\mu_j^{\;2}\right\rangle = \left(1/3\right)\mu^2$, $j = x,y,z$. Eqs. (34) then become

$$\frac{dn_1}{dt} = -\frac{1}{\hbar\Omega_{12}}\mathbf{E}\cdot\frac{d\mathbf{P}}{dt},$$ (35)

$$\frac{dn_2}{dt} = \frac{1}{\hbar\Omega_{12}}\mathbf{E}\cdot\frac{d\mathbf{P}}{dt},$$ (36)

$$\frac{d^2\mathbf{P}}{dt^2} = -\Omega_{12}^{\;2}\mathbf{P} + \frac{2}{3\hbar}\Omega_{12}\mu_{12}^{\;2}\mathbf{E}\left(n_1-n_2\right).$$ (37)a

This form of the Liouville-von Neumann equations, written in terms of the macroscopic populations and polarization, is identical in the absence of relaxation processes, to the kinetic equations (sometimes referred to as rate equations) used in the laser literature[278-280]). We note in passing that the more general form of Eq. (37) that does not assume isotropy

$$\frac{d^2\mathbf{P}}{dt^2} = -\Omega_{12}^{\;2}\mathbf{P} + \frac{2\Omega_{12}}{\hbar}\boldsymbol{\mu}_{12}\left(\mathbf{E}\cdot\boldsymbol{\mu}_{12}\right)\left(n_1-n_2\right)$$ (37)b

can be particularly useful for describing interfacial optical response. Eqs. (37) is the gateway for coupling the Maxwell's equations equation reference goes here to dynamics of the molecular system (more on this including practical numerical algorithms is in the next Section). Focusing for now on the one-dimensional case we note in the linear regime where $n_2 \ll n_1$ so that $n_1 \sim n_M$ Eq. (37)b implies the following expression for the electric susceptibility $\chi_e\left(\omega\right) = P\big/\left(\varepsilon_0 E\right)$

$$\chi_e = \frac{2\mu_{12}^2 n_M}{\hbar\varepsilon_0\Omega_{12}}\frac{\Omega_{12}^2}{\Omega_{12}^2 - \omega^2}.$$ (38)

To account for population relaxation and dephasing one usually introduces phenomenological decay constants in (31).[281] Here, however, different practices have been used. For the two-level molecular model under discussion, the standard scheme used in Bloch-type equations such as (33)

$$\frac{d\rho_{11}}{dt} = -i\frac{\boldsymbol{E}\boldsymbol{\mu}_{12}}{\hbar}\left(\rho_{12} - \rho_{12}^*\right) + \gamma_{21}\rho_{22},$$

$$\frac{d\rho_{12}}{dt} = i\left[\frac{\boldsymbol{E}\boldsymbol{\mu}_{12}}{\hbar}\left(\rho_{22} - \rho_{11}\right) + \Omega_{12}\rho_{12}\right] - \frac{\Gamma}{2}\rho_{12},$$

$$\frac{d\rho_{22}}{dt} = i\frac{\boldsymbol{E}\boldsymbol{\mu}_{12}}{\hbar}\left(\rho_{12} - \rho_{12}^*\right) - \gamma_{21}\rho_{22},$$

(39)

where $\gamma_{21}$ is the population relaxation rate, while the coherence $\rho_{12}$ decays with the rate $\dfrac{\Gamma}{2} = \dfrac{\gamma_{21}}{2} + \gamma_{\mathrm{d}}$, which takes into account pure dephasing rate $\gamma_{\mathrm{d}}$. These rates are related to the population relaxation time, $T_1$, decoherence time, $T_2$, and the pure dephasing time, $T_2^*$, in analogy with phenomenological formula from NMR spectroscopy[282]

$$\frac{1}{T_2} = \frac{1}{2T_1} + \frac{1}{T_2^*}.$$

(40)

Starting from (39) and following the procedure that leads to Eqs. (37)b now leads to the linear electric susceptibility in the form

$$\chi_e = \frac{2\mu_{12}^2 n_{\mathrm{M}}}{\hbar\varepsilon_0\Omega_{12}}\frac{\Omega_{12}^2}{\Omega_{12}^2 - \left(\omega - i\dfrac{\Gamma}{2}\right)^2}.$$

(41)

Another common phenomenological approach to account for relaxation, apart from the population relaxation rate $\gamma_{12}$ is to add a damping term to the polarization equation (37), hence replacing Eqs. (35)-(37) by[278-280]

$$\frac{dn_1}{dt} - \gamma_{21}n_2 = -\frac{1}{\hbar\Omega_{12}}\boldsymbol{E}\frac{d\boldsymbol{P}}{dt}$$

$$\frac{dn_2}{dt} + \gamma_{21}n_2 = \frac{1}{\hbar\Omega_{12}}\boldsymbol{E}\frac{d\boldsymbol{P}}{dt},$$

$$\frac{d^2\boldsymbol{P}}{dt^2} + \Gamma\frac{d\boldsymbol{P}}{dt} + \Omega_{12}^2\boldsymbol{P} = \frac{2}{\hbar}\Omega_{12}\mu_{12}^2\left(n_1 - n_2\right)\boldsymbol{E},$$

(42)

The linear electric susceptibility as follows from (42) reads

$$\chi_e = \frac{2\mu_{12}^2 n_\mathrm{M}}{\hbar \varepsilon_0 \Omega_{12}} \frac{\Omega_{12}^2}{\Omega_{12}^2 + i\Gamma\omega - \omega^2}, \tag{43}$$

which is different from (41). The susceptibility obtained from the Liouville-von Neumann equation differs from the conventional Lorentzian profile (43) at high dephasing rates resulting in noticeable red shift of the resonance from the molecular transition energy.

When molecules are located near a metal interface (a metal nanoparticle, arrays of thereof, arrays of holes, etc.) one should pay attention to the local field polarization, which is in general nonlinear and time dependent when an SPP mode is excited. This is illustrated with an example of a double-slit system as shown in Fig. 8. When excited by a resonant incident field the corresponding SPP mode is launched on both input and output sides of the film. Most importantly, the spatial regions near the slits that are displayed with red and blue colors correspond the local EM with horizontal and vertical polarization components. With molecules deposited on either side of the film this should be carefully taken into account.

The extension of the model based on rate equations to the case of exciton-plasmon materials with a local field being of arbitrary polarization is straightforward.[283] However the density matrix approach explicitly takes into account a particular configuration of the excited state wavefunction, which in turn determines the coupling with the local EM field. As frequently implemented, the Hamiltonian of a molecule is written in the basis of angular momentum states such that the dipole coupling with a local electric field usually given as a solution of Maxwell's equations in Cartesian coordinates must be written in spherical coordinates.[273, 284] Alternatively one may work in Cartesian coordinates for both EM field and molecular dynamics operating in the basis of $s$, $p$ orbitals. One can also invoke symmetry group algebra.[265]

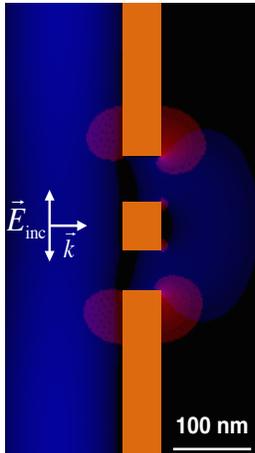

**Figure 8**. Two sub-diffraction slits in thin silver film. The system is excited by a resonant plane wave. Blue color shows instantaneous electromagnetic intensity associated with the vertical electric field component, while red indicates the induced horizontal component.

# 3. Numerical modeling

Due to the complex structure of modern plasmonic materials, it is only for a small number of geometries that one can, in principle, solve the Maxwell equations analytically in terms of the EM field modes. More often than not it is required to numerically integrate Maxwell's equations (28) for the specific structure and the given environment. Moreover, when dealing with hybrid systems comprising molecules near plasmonic interfaces such as nanoparticles, in particular under strong coupling conditions, a more involved theory requires numerical integration of the coupled Maxwell-Bloch equations. It is thus important to have access to spatiotemporal dynamics of EM field. There are various numerical methods available for numerical integration of Maxwell's equations on a grid.[285, 286] This part of the review discusses in depth one of many methods, namely finite-difference time-domain approach (FDTD). In addition to applications of FDTD in plasmonics we also provide several working FORTRAN codes, which can be found in Supplemental Material section. The supplied codes correspond to various applications discussed in details in Sections 2 and 4.

*FDTD and programing aspects*

Consider a system of Maxwell's equations equation reference goes here in an inhomogeneous space occupied by different dielectric components whose optical responses are determined by the local polarization currents. An attractive feature of solving Maxwell's equations directly instead of propagating a wave equation is a better numerical convergence since Eqs. (28) are first order equations unlike the wave equation. In a standard numerical approach one sets a numerical grid over the system and represents the curl operators in the Maxwell equations by finite differences at any grid point. When the dielectric function is location dependent such an approach requires an additional effort in order to account for proper boundary conditions at dielectric interfaces.[6] Such a task becomes very involved when dealing with complex geometries. It was shown, however, that specifically displacing the electric and magnetic field components one with respect to the other one can always satisfy the boundary conditions at every grid point. This proposal was put forward by Kane Yee in his seminal paper in 1966 [287] and later on was further developed by Allen Taflove.[288]

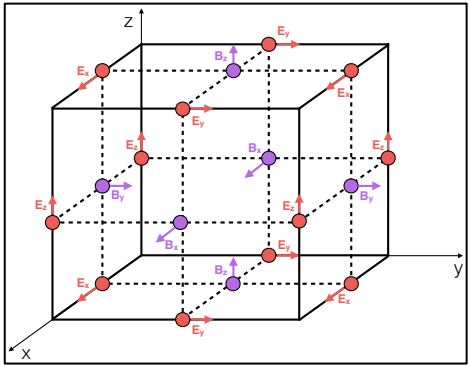

**Figure 9**. Yee's cell with the origin corresponding to the spatial point described by three integers $(i,j,k)=(x_i,y_j,z_k)$ located on the lattice. The side of the cube along each direction corresponds to the spatial step along this direction.

Fig. 9 illustrates the unit cell (so-called Yee's cell), which is at the heart of the FDTD scheme and the basis for all the codes that accompany this review. The lattice is defined by a set points, each given in terms of three integers $(i,j,k)$ corresponding to the discretized spatial coordinates $(x_i,y_j,z_k)$ an the origin is a point on this lattice. The cell's side in each direction is one spatial step. One can see that each component of the magnetic field is surrounded by the components of electric field that correspond to the curl operator. Similarly placing more Yee's cells next to each other results in electric field components surrounded by appropriate magnetic field. Each component of EM field is shifted by half a step in particular directions to accommodate the curl equations. As noted above, it is straightforward to show that the discretization of equations (28) using the grid based on Yee's cell leads to the fact that interface conditions are explicitly included in simulations. One would only need to properly discretize the dielectric function. We note that equivalently to the arrangement of EM field components in Fig. 9 we can interchange the electric and magnetic fields sublattices, a procedure frequently used in Ref. [288]. In our calculations (and the accompanying codes) we employ the arrangement shown in Fig. 9 as introduced by Sadiku[289] and Sullivan.[290] Once the curl equations are discretized in space one needs to propagate finite differences in time. This is frequently done using the leapfrog-time stepping technique, which first evaluates the magnetic field using Faraday's law and then based on the updated magnetic field the electric field is advanced in time. It should be noted that the opposite scheme, in which electric field is updated first, is equivalent to the conventional time propagation procedure. Since this approach is essentially an initial boundary value problem, knowledge of the EM fields at time $t = 0$ is required. Once the grid is initiated, *i.e.* an incident source begins to radiate, the field updates in time are done for all spatial positions on a grid. The most common practice in computational nano-optics is to use pulsed sources implementing either its so-called soft analogue or more commonly utilized the total field / scattered field approach (TFSF).[288] The latter uses the linear property of Maxwell's equations and splits the computation domain into two regions: a) total field region surrounding a nano-system, where both scattered and incident fields are evaluated, b) scattered region surrounding total field region, where only scattered fields are calculated. For systems with periodic boundaries, if one is interested only in a normal incidence geometry, a soft pulsed source can be easily implemented as shown in all periodic codes accompanied this review.

In one or two dimensions FDTD is relatively fast and does not necessarily require much of the computer memory. In contract, three-dimensional simulations are quite costly in terms of time and memory usage. However, owing to the finite differences discretization in FDTD one can utilize multiple processors. The supplied codes are written using the message-passing interface (MPI), which is widely implemented in, for example, the freely available Rocks Cluster package. The idea of parallel evaluation of two-dimensional FDTD equations is schematically depicted in Fig. 10. Here a total number of processors used is 8, each of which has its own unique number called *rank* that starts with 0 for the root processor. One can decompose FDTD grid onto 8 slices along *x*-axis as shown in the Figure 10, or vertically along *y*-axis – either decomposition leads to nearly identical speedup factors as long as *x* and *y* dimensions are the same. Clearly if a given problem requires elongated geometry the best way to choose a parallel grid is to slice the FDTD domain along shorter direction. The next step is to implement send and receive operations allowing neighbor processors to exchange data as schematically shown in Fig. 10.

When simulating EM wave propagation in open systems, the spatial grid must be terminated with particular boundary conditions ensuring that outgoing waves leaving the system do not artificially come back. A simple termination of the grid is equivalent to replacing boundaries of the simulation domain with perfect electric conductor. In this case EM radiation never leaves the system and results in unphysical field-matter interaction. Unlike in numerical integration of the Schrödinger equation where one propagates a wavepacket and can easily absorbs outgoing probability density flux via the exponential window function or the optical potential method,[291] absorbing boundary conditions for the Maxwell equations are more involved since the latter corresponds to the vector field and absorption must be taken care of at all possible scattering angles. A widely used approach is to surround the simulation domain by a set of unphysical material referred to as perfectly matched layers (PMLs), the transmission and reflection of which are adjust such that the incoming wave enters PMLs with the minimum reflection and maximum transmission and then attenuated inside PMLs. The cornerstone of this approach is related to different types of PML coefficients for different EM field components such that to minimize the reflection and maximize the transmission at all angles. The original idea, proposed by Bérenger,[292] was to split each component of EM field into two sub-components corresponding to two contributions from the curl operator and then apply PMLs separately for each sub-component. Since the original paper[293] several extensions to PML method were proposed. Of these, the easiest to implement but yet very powerful is the so-called convolution PML (CPML) technique.[294] It has a very important advantage over the split-component method: it efficiently absorbs evanescent fields without causing numerical instabilities - obviously a highly desirable feature in numerical plasmonics. Our codes are supplied with CPML absorbing boundaries. The particular CPML coefficients seen in the codes were carefully determined in numerous numerical experiments[288] and extensively tested on the plasmonic and exciton-plasmon systems considered here.

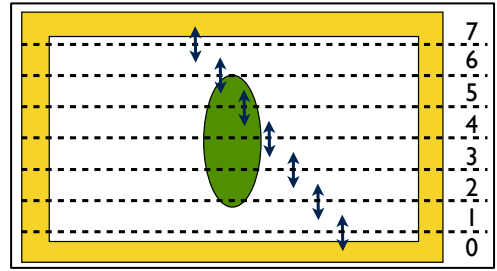

**Figure 10**. Schematics of parallelization of FDTD. Absorbing boundaries are shown as a yellow layer surrounding the grid. Dashed lines indicate the boundaries between processors. Vertical arrows indicate send/receive operations needed to calculate spatial derivatives at the boundaries.

An important technical point should be noted: The numerical stability of the FDTD propagation requires that the Courant condition [288]

$$\delta t < \frac{\delta x}{\sqrt{N} c}, \tag{44}$$

is satisfies, where $\delta t$ and $\delta x$ are the integration time step and the spatial resolution employed in the calculation, $N$ is the number of dimensions in the given problem and $c$ is the speed of light in vacuum. In all our codes the expression $\delta t = \delta x / (2c)$ is used to ensure numerical convergence irrespective of a number of dimensions involved.

Finally, the model dielectric functions for dispersive materials are given by equations (1)-(4) as functions of frequency. The FDTD solver requires the presentation of these functions in the time domain. The commonly used approach relies on properly defining new (auxiliary) equations capturing the dynamical behavior of the polarization current density (so-called the auxiliary differential equation method (ADE)[295] in spatial regions with the dispersion. To illustrate this method we consider the most general example of a material whose optical properties are described by the Drude-Lorentz function (2). The corresponding equations on the current density can be obtained from the relation between polarization, displacement, and electric field in the frequency domain

$$\boldsymbol{D} = \varepsilon_0 \varepsilon_\infty \boldsymbol{E} + \boldsymbol{P} = \varepsilon_0 \left( \varepsilon_\infty + \chi(\omega) \right) \boldsymbol{E}, \tag{45}$$

where the electric susceptibility is defined as $\chi = \dfrac{\varepsilon}{\varepsilon_0} - 1$. The contribution to the displacement current from the dispersive second term in (45) reads

$$\begin{aligned}
\boldsymbol{J}_{\mathrm{D}} &= i\omega\varepsilon_0 \left( -\frac{\Omega_p^2}{\omega^2 - i\Gamma\omega} \right) \boldsymbol{E}, \\
\boldsymbol{J}_{\mathrm{L}}^{(n)} &= i\omega\varepsilon_0 \left( \frac{\Delta\varepsilon_n \omega_n^2}{\omega_n^2 + i\gamma_n\omega - \omega^2} \right) \boldsymbol{E},
\end{aligned} \tag{46}$$

here the current density is split in two parts, $i.e.$ $\boldsymbol{J}_{\mathrm{D}}$ is the Drude current density and $\boldsymbol{J}_{\mathrm{L}}^{(n)}$ is the Lorentz current density associated with $n^{\mathrm{th}}$ pole in (2). Finally, the corresponding auxiliary differential equations on the current densities in the time domain are[296]

$$\begin{aligned}
\frac{\partial \boldsymbol{J}_{\mathrm{D}}}{\partial t} + \Gamma \boldsymbol{J}_{\mathrm{D}} &= \varepsilon_0 \Omega_p^2 \boldsymbol{E}, \\
\frac{\partial^2 \boldsymbol{J}_{\mathrm{L}}^{(n)}}{\partial t^2} + \gamma_n \frac{\partial \boldsymbol{J}_{\mathrm{L}}^{(n)}}{\partial t} + \omega_n^2 \boldsymbol{J}_{\mathrm{L}}^{(n)} &= \varepsilon_0 \Delta\varepsilon_n \omega_n^2 \frac{\partial \boldsymbol{E}}{\partial t}.
\end{aligned} \tag{47}$$

The equations (47) are then discretized in time and are propagated along with the Maxwell equations in spatial regions occupied by the dispersive material.

This relatively simple and yet powerful FDTD technique and its extensions has attracted considerable attention, resulting in several commercial packages, yet it is important to emphasize that the availability of home-built codes is often crucial. Although commercial packages offer a wide variety of options including automatic parallelization, they lack the versatility needed for advance reseseach, including many important features as discussed below. Despite the wide variety of commercial products, a cutting-edge research depends critically on capabilities that reach far beyond those of black-box commercial codes – for example, integrating Maxwell's equations on a grid with additional propagations of the Liouville-von Neumann equation, which is not included in any commercial packages. In passing, it should be mentioned that recently several research teams began to use simulations on graphical cards – graphical processor unit (GPU) FDTD.[288] Although GPU simulations can be very efficient compared to standard symmetric multi-processor (SMP) clusters, there is a fundamental limitation to speedups one can achieve using GPU. For example, to be able to perform simulations similar to some of those presented below one needs to use a graphical card with a memory larger than 32 Gb, which is not commercially available yet. In order to perform such simulations it is required to access RAM memory on a motherboard, which dramatically slows simulations and results in speedup factors of less than 1, i.e. computations are actually slower on several GPUs compared to a single processor.

*Numerical integration of coupled Maxwell-Bloch equations*

We now turn to the discussion of exciton-plasmon materials and numerical implementation of a Maxwell-Bloch integrator. The original idea of numerically integrating coupled Maxwell-Bloch equations was published by Ziolkowski, Arnold, and Godny in 1995.[297] Consider interaction of EM radiation with a set of two-level quantum emitters in one dimension. The dynamics of each emitter is governed by the Liouville-von Neumann equation (39) with the local electric field that has contributions from all other emitters and the incident radiation. Each emitter contributes to the total EM field via dipole radiation that is taken into account via corresponding polarization currents. The Maxwell's equations (28) are coupled to the quantum dynamics of the emitters, and the system of corresponding partial differential equations needs to be propagated in time. In their original paper Ziolkowski et al.[297] proposed the iterative procedure based on a predictor-corrector scheme to numerically integrate this system of equations. Such a procedure was also adopted in two[265] and three dimensions.[269] One starts iterations with initial conditions corresponding to emitters in the ground electronic state and the incident field is injected into the grid via TF/SF or by other means.[298] The solution vector, $U$, that contains all components of the EM field and the density matrix elements for each emitter, is updated according to

$$U^{(\text{new})} = U^{(\text{old})} + \delta t F\left(U^{(\text{old})}, U^{(\text{new})}\right),$$ (48)

where $F$ is the functional based on Maxwell-Liouville-von Neumann equations. It was shown that the proposed scheme converges after about 3-4 iterations for each time step. The convergence, however, is significantly diminished at high emitters' concentrations. It also becomes noticeably slower when applied to exciton-plasmon systems at resonant conditions[269] when emitter's transition energy is close to an SPP mode of the plasmonic system.

Bidégaray proposed an alternative numerical scheme similar to the ADE method that is used to numerically describe the optics of dispersive materials.[299] The premise is to split Maxwell's equations from the quantum dynamics by a half a time step. The resulting numerical procedure as outlined below does not require several iterations for each time step (which may become very costly when attempting to obtain steady-state solutions). The method is called a weakly coupled technique and was proven to be very efficient especially in two and three dimensions. The resulting integration procedure of the coupled Maxwell-Bloch equations can be outlined as follows:

1. In the regions occupied by quantum media Maxwell's equations are solved utilizing FDTD algorithm. First, magnetic field is updated according to Faraday's law. Next, using Ampere's law we update the electric field with the macroscopic polarization current density, which uses the density matrix at the previous time step. The EM field in the regions occupied by metal is updated according to the ADE method evaluated at the same time step as the density matrix components. With the knowledge of electric field components (stored in memory at two time steps) we update the density matrix at each spatial point on the grid according to (39) using a suitable numerical technique for the integration of a system of ordinary differential equations. Usually the fourth order Runge-Kutta scheme is sufficient.

2. With the knowledge of electric field components and updated density matrix we calculate the macroscopic polarization current at each grid point.

Clearly the weakly coupled method is not limited to a particular equation describing quantum dynamics. One can use this method and couple Maxwell's equations to a set of rate equations (42), for instance.

*Notes on linear and nonlinear simulations*

The FDTD algorithm is a time domain method that requires extra-effort when calculating steady-state solutions of the Maxwell equations or a particular spectral response such as a scattering cross-section. In the linear regime, when optical properties of the system under consideration are independent from the incident field, it is possible to obtain a spectral response within a single FDTD run. The method called the short pulse method (SPM) was first introduced in Ref[300]. The system is excited by a short pulse of the bandwidth that covers the spectral range of interest. Corresponding equations are propagated for extended period of time such that the EM fields decay due to intrinsic losses in the system and radiation to the far field. During the time iterations the discrete Fourier transform of EM field components is evaluated on-the-fly. After reaching the end of the time propagation the calculated Fourier components are weighted with respect to the incident pulse contributions at a given frequency (that includes both an amplitude and a phase). This method thus yields spectra in the frequency domain within a single FDTD run.

Because the SPM relies on matching Fourier-frequency components of the output signal with the corresponding components of the incoming pulse, this method cannot be used for processes involving inelastic scattering or nonlinear response. A common example of the former type is the response of a molecular system exhibiting scattering or fluorescence at frequencies different from the absorption peak. Here the conventional continuous wave method (CWM) must be applied. One drives the system using CW pump at a given frequency $\omega_{in}$, steady-state solutions are obtained, a given observable is calculated, and the process is repeated for another $\omega_{in}$ (more on this subject below).

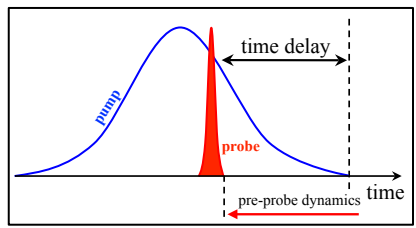

**Figure 11**. Illustration of the pump-probe simulations.

Let us now consider an example of nonlinear spectroscopy where materials are subject to a high intense fs laser pulse pump. Then after some controllable time delay the low intense probe is applied measuring a particular spectral observable such as reflection or transmission, for instance. The goal of this technique is to be able to track the dynamics of a system driven by the pump recording changes of the observable compared to a linear case (*i.e.* when the pump is not applied).

The key challenge in modeling nonlinear dynamics using the pump-probe pulse sequence is to disentangle signals caused by the strong pump and the weak probe. When a system comprised of optically coupled emitters is excited by intense pump pulse, it exhibits polarization oscillations lasting long after the pump is gone. Consequently, when the system is probed by a weak probe, one observes an undesired high intensity signal at the pump frequency caused by induced polarization oscillations. Experimentally, such oscillations may be filtered out so as not to interfere with the probe in the far field. In simulations, these unwanted oscillations must be handled carefully because they may interfere with the signal produced by the probe. The efficient computational method to simulate transient spectroscopy experiments was proposed in Ref. [301]. The idea behind this method is illustrated in Fig. 11. A high intensity pump drives the system then the weak probe is sent after a given time delay. The corresponding equations of motion are propagated in time and space with the driving pump up until the probe pulse "arrives". At that time the density matrix elements are recorded at all grid points where quantum emitters are located. These data are used as initial conditions for simulation of the probe interaction with the sample. This method guarantees that undesired high amplitude oscillations are absent when one probes the system and that the probe does not alter the optical response of the system. The method is however approximate as it disregards pump-probe coherences that may affect the actual experimental system.

To illustrate how such a technique works we consider a pump-probe experiment applied to a thin one-dimensional layer comprising interacting two-level molecules with molecular resonance at 1.6 eV. Our interest is to apply a pump-probe pulse sequence and evaluate the absorption as a function of the incident photon frequency and the pump-probe delay for different pump intensities. Fig. 12 shows results of simulations based on the numerical technique described above. The time envelope of the pump pulse is taken in the form[282]

$$\int_{-\infty}^{+\infty} F(t) dt = n\pi \frac{\hbar}{2\mu_{12}E_0},$$

(49)

where $n = \frac{1}{2}, 1, \frac{3}{2}, ...$ and $E_0$ is the pump amplitude. We start with a pump corresponding to the $\pi/2$ pulse

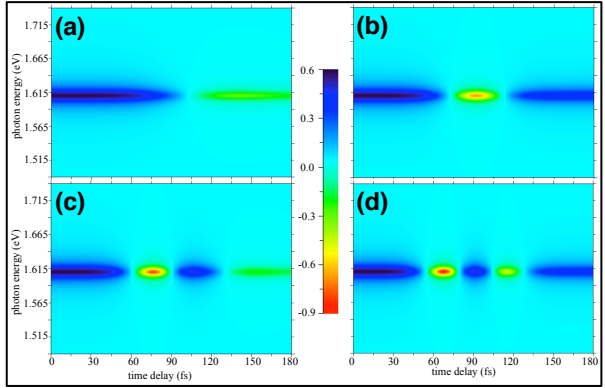

**Figure 12.** Absorption in the pump-probe simulations as a function of the incident photon energy and time delay. Panels (a) through (d) show how absorption changes with time when $\pi/2$ (a), $\pi$ (b), $3\pi/2$ (c), and $2\pi$ pumps are applied. The pump is 180 fs long.

($n = \frac{1}{2}$) that completely inverts a single molecule, transferring population initially located in the ground state to the excited state. It is seen that the system exhibits positive absorption during the first half of the pump (even slightly longer than 90 fs) and then the sign of the absorption changes as molecules are in their excited state. The probe pulse triggers stimulated emission thus resulting in a negative absorption. Other panels in Fig. 12 demonstrate the results for the pumps with $\pi$, $3\pi/2$, and $2\pi$ envelopes.

As anticipated, the molecules undergo Rabi oscillations that depend on the pump pulse area. However, the Rabi cycling is not perfect as it would be if the molecules were independent and not interacting. Each subsequent oscillation of the system from the ground state to the excited state is less pronounced due to decoherence.

*Semiclassical treatment of fluorescence*

The semiclassical model based on coupled Maxwell-Bloch equations cannot account for the phenomenon of spontaneous emission since the latter is a purely quantum effect associated with the quantum nature of the radiation field. It has been demonstrated that the effect of spontaneous emission can be partially accounted for by imposing a classical stochastic field on the system.[266] Obviously, classical electromagnetic noise cannot mimic vacuum fluctuations; in particular, it can induce excitation of ground-state molecules while vacuum fluctuations can lead only to radiative damping of excited molecules. One can use this trick to study time evolution in a system, which is initially inverted; that is, all molecules are in the excited state. In this case, the induction of emission by the EM noise will soon lead to a dominant signal of induced emission that does not depend much on the nature of that noise, provided the latter is weak enough.[270] This method has been successfully used to simulate superradiance and gain within the FDTD approach.[274] However, this method cannot be used to generate fluorescence in a system of mostly ground-state molecules, where the main effect of such noise will be to induce unphysical molecular excitation (as if a given molecule being embedded in the "noise" field undergoes the unphysical process of absorbing its own spontaneously emitted photon).

It is therefore highly desired to have an algorithm that: 1) still relies on Maxwell's equations to evaluate spatiotemporal dynamics of electromagnetic radiation; 2) does not include nonphysical excitation of relaxed molecules by "their own spontaneous photons"; 3) produces EM field due to spontaneous emission that has a proper time correlation function and dependence on damping parameters (radiationless transition rate, pure dephasing rate). It was recently proposed [302] that this could be accomplished in a straight manner by inserting a dipole source into Ampere's law with specific time characteristics. Let us briefly overview the method. The idea is that in addition to the classical scattering by the molecule, the molecule itself also acts as a source of spontaneous radiation, which has the properties of a random electromagnetic field. The latter then is added as a dipole source to the Ampere law. The time dependence of the stochastic component of the field can be deduced by considering properties of electromagnetic field produced by a single molecule at the steady state. Upon careful derivation of the corresponding field correlation function the stochastic field due to spontaneous emission is obtained[302] by employing the stochastic dipole

$$d_R(t) = \sum_{m=1}^{N} \sqrt{d_m} A_m \cos(\omega_m t + \varphi_m),$$ (50)

where $A_m$ and $\varphi_m$ are the Gaussian random number and the random phase between 0 and $2\pi$, respectively, the sum is taken over $N$ oscillators ($N$ needs to be chosen such that the numerical convergence is achieved when a particular physical observable is calculated). The normalization factor $d_m$ is defined as

$$d_m = \frac{(\Delta\omega)c\hbar\Gamma\Gamma_R}{\mu_0\Omega_{12}^3}\left|\rho_{22} - |\rho_{12}|^2\right|\frac{N_{\text{total}}}{\left(\Omega_{12} - \omega_m\right)^2 + \left(\frac{\Gamma}{2}\right)^2},$$ (51)

here $\Gamma_R$ is the radiative lifetime of the excited state, $\Gamma = \Gamma_R + \gamma_{21}$, $\Delta\omega$ is the spectral resolution of the sum in (50) sampled around the molecular transition frequency, $\Omega_{12}$, excited state population, $\rho_{22}$, and the coherence, $\rho_{12}$, are included in (51) to account for the pure dephasing. $N_{\text{total}}$ is the total number of molecules at given grid point where the stochastic dipole is generated. The dipole (50) is inserted into the Ampere law.

Fig. 13 illustrates the aforementioned approach with the example of a single two-level molecule driven by an off-resonant CW radiation. The molecular transition frequency is set at 3 eV and the laser field drives the molecule at 2.85 eV. The intensity of the radiation produced by an oscillating pointwise dipole at steady state is proportional to the square of the dipole moment. The molecule is initially in the ground state and is driven by a low intensity incident field such that the population of the excited state is always small. The corresponding dynamics is evaluated using Bloch equations for the density matrix. With the fluorescence present, the scattering of EM radiation is no longer elastic and for a given CW frequency there is a distribution of scattered frequencies. The total emission/scattering signal is thus evaluated as an integral over all scattered frequencies, however it is informative to separate the signals due to spontaneous emission (50) and coherent radiation produced by off-diagonal elements of the density matrix. Fig. 13 shows the linear dependence of the fluorescence on the pure dephasing rate at low dephasing. One can also see that the fluorescence becomes dominant when the dephasing becomes faster, exceeding $10^{-2}$ eV for the present choice of parameters.

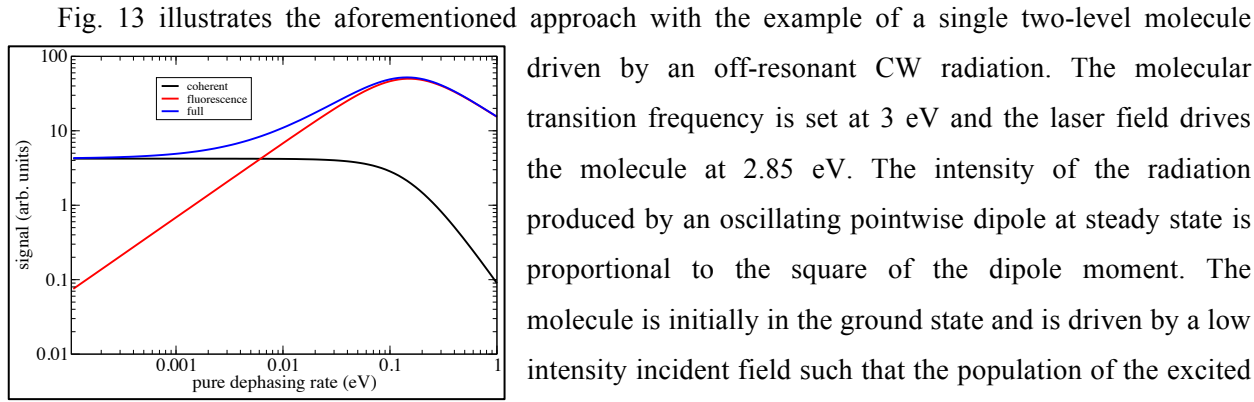

**Figure 13**. Signal as a function of the pure dephasing rate. Black line shows the coherent signal, the fluorescence is shown as a red line, and the complete signal is shown as a blue line.

## 4. Applications

The semiclassical theory described in previous sections inherently takes into account all electromagnetically induced interactions in an exciton-plasmon system, limited only by the classical description of the electromagnetic field and the mean field approximation for the molecular response. It can thus describe a vast body of various phenomena pertaining to the linear and nonlinear optical response of the many-molecule system interacting with plasmonic nanostructures, including collective effects. This section discusses several intriguing applications of the theory to understand both linear and nonlinear optical properties of exciton-plasmon nanomaterials.

*Dipole induced electromagnetic transparency*

Our first example is the problem of reflection/transmission/absorption of low intense radiation by a one-dimensional layer comprised of interacting two-level molecules. In a series of papers Prasad and Glauber attempted analytically to describe the rich physics of this problem by proposing a new model called polarium.[94-97] The latter consist of interacting randomly oriented two-level emitters. It was shown that the system exhibits a set of multiple resonances near the molecular transition frequency. The reflection was shown to have a flat plateau reaching unity at high molecular concentration. The theoretical analysis , however, relied on a pure mathematical description with a little physical insight.

Later on the semiclassical theory based on Maxwell-Bloch equations was used to re-visit the problem of scattering of EM radiation by thin molecular films.[303] The complete theory in this paper, in addition to the Maxwell-Bloch equations, relied on somewhat artificial term included as a local field correction in Bloch equations, i.e. the so-called Lorentz-Lorenz (LL) correction.[262] Such a term allows one to work out a simple analytical model capable of explaining the observed scattering in a straightforward manner. The LL correction remains however a controversial issue.[304] A simple counterargument is as follows. Focusing on the scattering of EM radiation by a thin molecular film comprised of two-level emitters, it is obvious that at low molecular densities the transmission at normal incidence has a minimum at the molecular transition frequency, which is indeed observed in the Maxwell-Bloch simulation. Let us now assume that we keep increasing the spatial resolution when integrating the corresponding Maxwell-Bloch equations. At relatively low molecular densities the best spatial resolution we can have is such that each grid point is occupied by a single molecule inside the molecular layer. Since we solve Maxwell-Bloch equations exactly all electromagnetically induced interactions are included. Thus we conclude that the LL shift of the resonance exhibited by the transmission should be observed without artificially including the

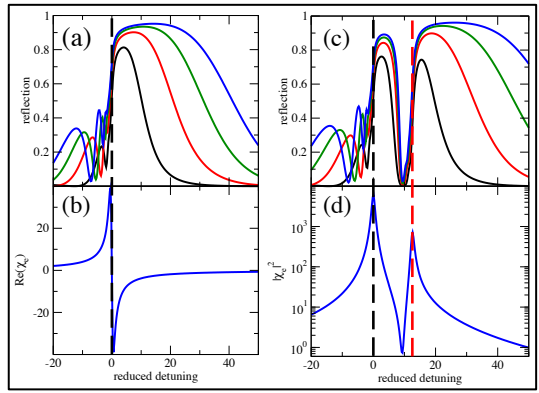

LL local field correction. Indeed, it was shown in Ref. [304] by directly including contributions from tens of millions of pointwise dipoles that there is no observed LL shift of the molecular resonance. (It should be noted however that an experimental evidence of the LL shift has been reported.[305, 306])

While the question of a proper microscopic theory that would correctly account for the LL shift remains open, we turn to the discussion of the main results reported in Refs. [303, 307]. As noted above, the interpretation advanced in these paper relies on the LL shift. Here we note that numerical simulations based on direct integration of Maxwell-Bloch equations support the main findings of the Refs. [303, 307] without explicitly invoking such shift.

**Figure 14**. Linear optical response of thin molecular films. Panels (a) and (c) show reflection spectra as functions of reduced detuning for films comprised of two-level molecules and three-level molecules, respectively. Corresponding susceptibilities are shown in panels (b) and (d). Vertical dashed lines indicate molecular transition frequencies.

Fig. 14 shows the results of such simulations of the linear optical properties of thin molecular films. The reduced detuning is defined as $\delta = \frac{2}{\Gamma}\left(\omega - \Omega_{12}\right)$. Panel (a) shows how the reflection spectrum for interacting two-level molecules varies with molecular concentration. The reflection becomes noticeably wider at high concentrations with the maximum shifting towards higher energies (very surprising result as one would expect to see the red shift predicted by the LL model). This is consistent with earlier results reported in Refs. [94, 95]. In contrast, the transmission spectra exhibit wide minima suggesting that the film becomes opaque. An elegant explanation of the obvious plateau exhibited by the reflection at high concentrations was proposed in Ref. [303]. We note that the real part of susceptibility is negative in the region of high reflection as seen in Fig. 14b, which means that molecules oscillate out-of-phase with respect to the incident field. The medium is then characterized by a collective dipole excitation that cancels out transmission and enhances reflection over a very large window around the transition frequency much like conductive electrons in metal. This phenomenon dominates in the case of high densities, where the dipoles coherently cooperate to prevent penetration of the incident radiation in the film. Resonant modes observed at a negative detuning are conventional Fabry-Pérot resonances.[307]

Completely different physical picture is observed when two-level molecules are replaced by emitters characterized by a V-type three levels. Fig. 14c shows reflection spectra for interacting three-level molecules at different concentrations. The widening of the reflection window and blue shift of the resonance is still seen. However, the major difference from the previous example is a clear wide reflection minimum that appears at the energy somewhat between two molecular transitions. The corresponding transmission (not shown) exhibits a maximum exactly at the minimum of the reflection. The system thus displays the transmission window in otherwise nearly completely opaque frequency range. The material parameters such as transition dipoles for two molecular transitions, their frequencies, the thickness of the film, and molecular concentration control the position of the transparency window.

In order to understand the physics behind this phenomenon it is informative to examine the absolute value of the susceptibility as shown in Fig. 14d. The susceptibility exhibits two resonances with different profiles. The low energy transition has a simple Lorentzian shape, while the high-energy transition shows up as a clear Fano resonance with a minimum located at the same frequency as the transparency resonance. This suggests that the interference between two dipole transitions plays a key role in this effect. There is a frequency range where two molecular transitions overlap and the corresponding contributions from the two types of dipoles add up in the macroscopic polarization of the medium. In addition, in this frequency range the two dipoles oscillate in opposite directions (out of phase). This comes from the fact that one type of emitters is blue detuned while the other is red detuned, leading to opposite signs of their susceptibilities. This transparency phenomenon is reminiscent of electromagnetically induced transparency (EIT).[308] Compared to EIT, however, the strong coupling induced by the pump laser is replaced by strong dipole-dipole interactions. This effect is a linear phenomenon and does not require intense laser radiation. By the analogy to EIT this effect was named Dipole-Induced Electromagnetic Transparency (DIET). DIET results in slow light,[307] although the group velocity estimated for interacting [85]Rb atoms reaches as low as 10 m/s, which is relatively high compared to EIT results.

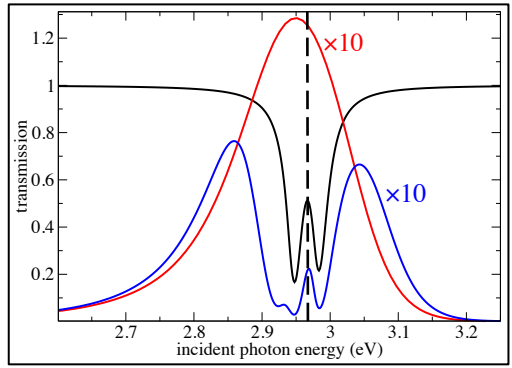

**Figure 15**. Three-dimensional DIET in exciton-plasmon systems. Transmission for a stand-alone molecular film is shown as a black like. Transmission for the bare hole array is indicated as a red line. Transmission calculated for the exciton-plasmon system is shown as blue line.

DIET is highly robust to material parameters. We performed three-dimensional simulations specifically for this review in order to illustrate that DIET can be observed in exciton-plasmon systems. The system under consideration is a conventional periodic array of holes in a silver film. The geometrical parameters are: periodicity is 320 nm, film thickness is 200 nm, the hole diameter is 140 nm, and the molecular film thickness is 100 nm. The molecular is placed directly on top of the array at the incident side. Corresponding surface plasmon-polariton resonance of the bare hole array is 2.95 eV. The molecular transition energies for are set at 2.95 eV (directly at the plasmon mode to achieve strong coupling) and at 2.985 eV. At the molecular number density of $4 \times 10^{25}$ m$^{-3}$ a stand-alone molecular layer exhibits DIET as seen in Fig. 15 as a maximum in the transmission. The corresponding hybrid system comprised of the periodic array of holes and molecular layer clearly shows very same transmission resonance located inside Rabi splitting. We note that due to strong coupling of the low energy molecular transition with the plasmonic mode there are two expected hybrid states seen in the transmission at 2.86 eV and 3.04 eV.

*Linear optics of excitonic clusters*

We now turn to the discussion of linear optical properties of ensembles of interacting molecules in two and three dimensions. Our goal is to establish general physical understanding of how such systems react to external optical excitation. Once equipped with such knowledge we shall discuss how these properties are altered when molecules are resonantly coupled to plasmonic systems.

As an example that illustrates how excitonic systems scatter electromagnetic radiation we consider an ensemble of two-level molecules uniformly distributed in a subdiffraction volume in the form of a circle (two dimensions) and a sphere (three dimensions).[273] Such a problem was first considered in a series of works by Prasad and Glauber,[94-97] where the framework of polarium model (see 4.1) was established. It was demonstrated that excitonic clusters might in principle exhibit a set of narrow resonances provided that the pure dephasing rate is negligibly small. Furthermore such resonances are always seen in the background of a wide mode, properties of which are attributed to the collective response of molecules to the incident field. Such a collective mode (Prasad and Glauber referred to this mode as a superradiant one) has unique characteristics that vary significantly with the molecular transition dipole and concentration. We note that such characteristics, which we shall discuss below, were recently experimentally verified.[309] It is also interesting to note that the superradiant mode in excitonic systems is observed only in two- or three-dimensional geometries. Fig. 16 shows results obtained for two- (panel (a)) and three-dimensional (panel (b)) clusters with a radius of 20 nm comprised of two-level molecules with a resonant energy of 3 eV. The scattering intensity clearly exhibits two resonances. The first mode is a relatively weak response slightly below the molecular transition frequency. The second is a strong and broad peak at a higher frequency that moves to the blue at larger molecular concentrations. We note that the intensity of the high-frequency mode scales as a square of the molecular density, *i.e.* $n_M^2$, suggesting a possible collective nature of this mode. It should be noted that this collective resonance is noticeably wider than the molecular transition resonance. Moreover, the result for increasing of the transition dipole, $d_{12}$, is qualitatively similar to that obtained with increasing molecular density. It is seen that the latter results in a blueshift of the collective mode and in a redshift of the lower-frequency mode.[273] The former shows a quadratic dependence of the resonant frequency on the dipole moment.

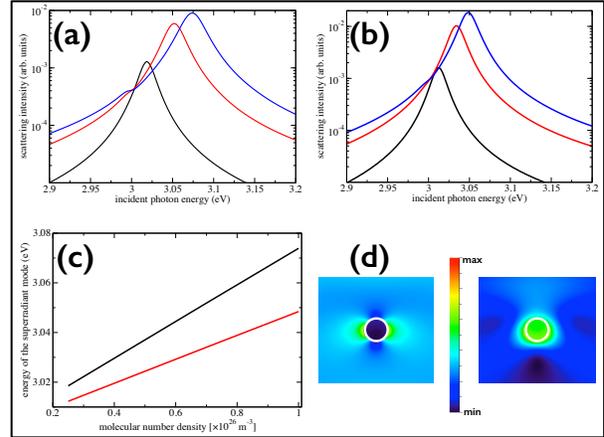

**Figure 16.** Scattering by two- and three-dimensional excitonic clusters. Panels (a) and (b) show scattering intensity as a function of the incident photon energy for two-dimensional (a) and three-dimensional (b) clusters at different densities: black line is for $2.5 \times 10^{25}$ m$^{-3}$, red line is for $7 \times 10^{25}$ m$^{-3}$, and blue line is for $10^{26}$ m$^{-3}$. Panel (c) shows the energy of the superradiant mode for two-dimensional (black line) and three-dimensional (red line) clusters as a function of the molecular density. Panel (d) illustrates intensity distributions for the surface mode (left) calculated at the molecular transition energy 3 eV and the superradiant mode (right) at 3.07 eV.

It has been shown numerically[97, 273, 310] that the width of the cooperative emission mode and its resonant frequency scale as

$$\begin{aligned}\Gamma_{SR} &\sim n_M \Gamma, \\ \omega_{SR} &\sim n_M d_{12}^2.\end{aligned} \tag{52}$$

The results based on direct numerical integration of Maxwell-Bloch equations follow the same dependence of the superradiant mode seen in Fig. 16 on the number density and the individual molecular decay rate. It is also informative to explore the spatial dependence of electromagnetic intensity associated with two observed resonances. Fig. 16d shows intensity distributions calculated using steady state solutions of Maxwell-Bloch equations at two frequencies. Clearly the electromagnetic intensity at the lower resonant frequency is mainly localized on the surface of the cluster exhibiting a dipole radiation pattern similar to the electromagnetic intensity distributions seen at the plasmon resonance for a single metal nanoparticle. The superradiant mode is uniformly distributed over the entire volume of the particle with all molecules coherently participating in the radiation process.

*Collective resonances in hybrid systems*

Systems comprised of metallic nanostructured arrays are particularly attractive for interaction with photonic molecular excited states, as the plasmonic resonances may be tuned by geometrical factors (i.e. the periodicity of arrays) to exactly match the molecular frequencies. As was pointed out in previous sections, when the coupling strength between the local field driven by plasmon oscillations and molecules exceeds all damping rates of the system the strong coupling regime is achieved. The strong coupling manifests itself as avoided crossing of the original non-interacting modes. The newly formed polaritons are energetically separated by the Rabi splitting, which reflects the coupling strength. Various experiments and theoretical calculations of a different complexity (ranging from simple two coupled oscillator models[177] coupled to fully quantum calculations[160]) have been performed to describe such a phenomenon. Most of the experiments, however, have been performed at relatively low molecular concentrations thus not capturing possible collective exciton resonances and their interaction with plasmonic modes.

It was recently demonstrated[273] that core-shell nanoparticles with a metallic core and a molecular shell resonant with the core localized plasmon mode may exhibit not two but three resonances depending on molecular density. The presence of the upper and lower polaritonic modes due to the hybridization of the plasmon mode and individual molecular resonance is, of course, expected. In addition to this observation the third mode seen at relatively high molecular concentrations is also observed in calculations.[273, 311] The physical nature of the third mode is related to the superradiant resonance discussed in the previous section since the mode's width and scattering intensity depend on the density, transition dipole, and shell's thickness similarly to the superradiant mode. Fig. 17a explores the density dependence of the collective resonance in the system comprised of a silver core and a resonant molecular shell. The Rabi splitting is seen at low concentrations. At higher concentrations the Rabi splitting increases and the third resonance appears near the molecular transition energy. Note that the position of the third mode shifts to higher energies with increasing concentration as in the case of the superradiant mode seen in Fig. 16.

The collective exciton resonance has been discussed in the context of periodic nanostructured arrays of slits,[192] where it was demonstrated that at high molecular concentrations the transmission exhibits a three-peaked structure similar to the example of the core-shell particle. It was shown that when the in-plane k-vector varies and the plasmon resonant energy sweeps through the molecular transition the avoided crossing for upper/lower polaritonic branches is observed. When the molecular density is high the third mode appearing in the transmission spectra is nearly dispersionless suggesting that indeed the third resonance is of collective nature mediated by the strong molecule-molecule interactions. The appearance of the collective mode in periodic systems leads to a large repulsion of the upper polariton and in some cases to enhanced transmission at specific frequencies. To provide a clear test of the collective nature of the third mode, calculations with molecule-molecule interaction artificially turned off were performed in Ref. [192]. No other changes were introduced meaning that plasmon-molecule coupling was still presented. Technically this was accomplished by replacing the molecular layer with a single two-level system occupying the same volume as molecular layer in which the local electric was assumed to be spatially uniform, thereby eliminating any internal dipole coupling, while maintaining all other interactions intact. The dipole-dipole interactions, which arise in the model based on Maxwell-Bloch equations from the field-induced interactions between individual molecules, are therefore absent. It was clearly demonstrated that indeed when molecular dipole-dipole interactions were excluded, the value of the Rabi splitting was smaller and the third mode was not observed. Even at significantly higher molecular densities, when the Rabi splitting exceeded 200 meV, the third peak was absent. Fig. 17b shows transmission spectra for a periodic array of slits with a 20 nm thin molecular layer placed on the input side of the array. The Rabi splitting increases with increasing concentration and the third resonance is observed near the molecular transition energy. The energy of the collective mode increases with increasing concentration similar to the core-shell system. We note that indications for this mode were found in several early experimental reports[110, 259], but its nature and origin have not been understood at the time.

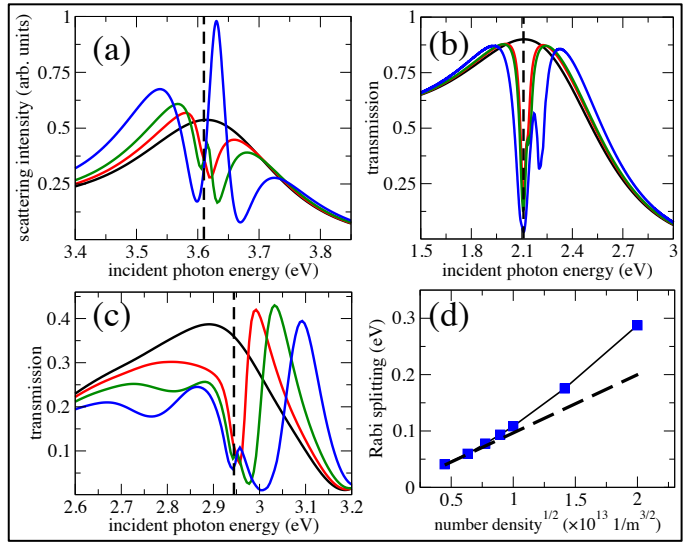

**Figure 17.** Collective exciton resonances in hybrid systems. Panel (a) shows the scattering intensity as a function of the incident photon energy for a core-shell nanoparticle with a silver core of 20 nm in diameter and excitonic 10 nm thick shell. The black line shows the intensity scattered by the bare silver core; the red line represents the scattering by the core-shell particles with the density of molecules of $8 \times 10^{24}$ m$^{-3}$; green line is for the density of $2 \times 10^{25}$ m$^{-3}$; the blue line is for the density of $6 \times 10^{25}$ m$^{-3}$. The vertical dashed lines in all panels indicate the molecular transition energy. Panel (b) shows the transmission through the periodic array of slits in a silver film with a period of 300 nm, with a 20 nm thin layer of molecules deposited on top of the film. The black line shows the scattering from the bare slit array; red, green and blue lines are similar results obtained in the presence of molecular films for the molecular densities of $5 \times 10^{25}$ m$^{-3}$, $7 \times 10^{25}$ m$^{-3}$ and $2 \times 10^{26}$ m$^{-3}$, respectively. Panel (c) shows transmission through the hybrid opal array with the period of 320 nm covered by a 20 nm thin layer of resonant molecules. The black line shows the transmission through the bare opal array and the red, green and bluelines are similar signals obtained with molecular layers of densities of $8 \times 10^{25}$ m$^{-3}$, $2 \times 10^{26}$ m$^{-3}$ and $4 \times 10^{26}$ m$^{-3}$ respectively. Panel (d) shows the Rabi splitting for the opal array in (c) as a function of square root of the number density. The deviation form linear dependence (the dashed line) is evident.

Recently experiments were performed on plasmonic opal arrays with J-aggregates deposited on metal surface with the molecular density being a controllable parameter.[312] The idea is to manipulate strong coupling between the Bragg-plasmon mode[313] supported by an opal array and J-aggregates dispersed on the input side of the hybrid structure. The collective exciton resonance was experimentally observed for the first time in these experiments. The transition from a conventional strong coupling regime exhibiting the usual upper and lower polaritonic branches to a more complex regime was clearly demonstrated. The third nondispersive mode was seen in reflection spectra, as the concentration of J-aggregates was increased. Rigorous numerical simulations based on three-dimensional coupled Maxwell-rate equations qualitatively support these experimental results. Figure 17c shows transmission spectra for the opal array at different molecular concentrations. One can see the appearance of the collective resonance reminiscent to the third mode we already discussed for other geometries. Furthermore it is clear from Figure d that the Rabi splitting deviates from the conventional $\sqrt{n_M}$ dependence at high densities when the collective mode is seen in spectra.

A simple semiclassical analytical model explaining the physical features of the collective mode was proposed.[312] Let us consider N identical molecules interacting with one another and with the background surface plasmon-polariton field. The Hamiltonian of such a system reads

$$\hat{H} = \begin{pmatrix} \varepsilon_{12} & C & C & \ldots & \Delta \\ C & \varepsilon_{12} & C & \ldots & \Delta \\ C & C & \varepsilon_{12} & \ldots & \Delta \\ \ldots & \ldots & \ldots & \ldots & \ldots \\ \Delta & \Delta & \Delta & \ldots & \varepsilon_{SPP} \end{pmatrix}, \tag{53}$$

where $\varepsilon_{12} = \hbar\Omega_{12}$ is the molecular transition energy, $\varepsilon_{SPP}$ is the energy of the corresponding plasmon mode, $\Delta$ is the molecule-plasmon coupling (deduced from the Rabi splitting), and $C$ is the phenomenological constant describing the coupling between molecules. This model assumes that molecules are interacting with one another with the same strength. This assumption obviously holds as long as the size of a system is smaller than the corresponding wavelength, and would neglect retardation effects if applied to the systems used in Fig. 17. Qualitatively, however, the proposed model predicts the correct behavior of the third resonance while if one needs to describe it quantitatively, the numerical integration of Maxwell-Bloch equations is required.

The Hamiltonian (53) is easily diagonalized leading to the following set of eigenenergies

$$E_n = \varepsilon_{12} - C, n = 1, \dots, N-2,$$
$$E_{N-1,N} = \frac{(N-1)C + \varepsilon_{12} + \varepsilon_{\text{SPP}}}{2} \pm \sqrt{\left(\left(N-1\right)C + \varepsilon_{12} - \varepsilon_{\text{SPP}}\right)^2 + 4N\Delta^2}, \quad (54)$$

here the first root is $(N-2)$-degenerate and the other two represent the upper and lower polaritonic branches although noticeably altered by molecule-molecule interactions. The Rabi splitting, $\Delta E$, defined as the energy difference between upper and lower polaritonic branches at the zero-detuning condition $\varepsilon_{12} = \varepsilon_{\text{SPP}}$ is

$$\Delta E = \sqrt{\left(N-1\right)^2 C^2 + 4N\Delta^2}. \quad (55)$$

At low molecular concentrations, as the molecule-molecule interaction energy $C$ is negligible, the Rabi splitting scales as $\sqrt{N}$ as in the conventional model.[108] Moreover, the scaling of the Rabi splitting with respect to the molecular transition dipole is linear since the molecule-plasmon coupling constant $\Delta$ also scales linearly with the dipole. This model thus contains the conventional model based of two coupled oscillators as a limiting case. Let us now examine the behavior of the Rabi splitting at high molecular concentrations. The scaling of the Rabi splitting with molecular density at high $N$ is

$$\Delta E \approx NC. \quad (56)$$

The clear deviation of the Rabi splitting from the expected $\sqrt{N}$-dependence is observed. This has also been observed in numerical simulations performed for all systems discussed in Fig. 17.

It is of interest to determine how the coupling constant $C$ scales with molecular parameters. This can be done by fitting the results (54)-(56) to those obtained by numerical simulations, and such fitting was done by simulating systems with varying molecular transition dipole moment, $d_{12}$, and molecular density at the resonant conditions corresponding to opal arrays, arrays of slits, and periodic arrays of holes (for both rectangular and honey-comb arrangements of holes). From all these simulations a nearly ideal quadratic dependence of the coupling $C$ on the dipole moment was observed. Moreover, a clear linear scaling with the molecular density at intermediate densities was also seen. Using a simple argument of the mean inter-particle distance scaling with the number density $\langle R \rangle \sim 1/N^{1/3}$, we arrive at the following expression for the coupling constant $C$

$$C \sim N d_{12}^2 \sim \frac{d_{12}^2}{\langle R \rangle^3}. \qquad (57)$$

The coupling constant evidently has a form of the potential energy of a dipole in the field of another identical dipole thus confirming earlier prediction[192] that the observed third mode corresponds to the collective exciton resonance. We note that the scaling of the collective mode with molecular concentration and the dipole moment is the same as in (52). Moreover this expression combined with (56) results in a clear quadratic dependence of the Rabi splitting on the molecular density. This model is independent from the geometry suggesting that the results discussed here should be observed in other hybrid systems.

We note that the collective mode discussed in this section does not require a presence of the plasmon field. All attributes of the collective nature of this mode pointed out above such as quadratic dependence on the transition dipole were found in excitonic clusters as well.[273] One can argue that for a given molecular concentration higher exciton-plasmon coupling (i.e. higher values of the Rabi splitting) may in principle lead to an observation of the collective mode as long as the damping at the molecular transition energy is lower than the strength of molecule-molecule interaction.

*Nonlinear molecular plasmonics*

The presented semiclassical theory can also be applied to describe nonlinear optical phenomena since the molecular part of the theory has no restriction on how high the local electric field amplitude may be. We end this review by providing several representative examples of hybrid systems exposed to strong external laser radiation.

The way the pure dephasing is introduced in the theory is through simple damping terms in corresponding equations governing quantum dynamics. This corresponds to homogeneous broadening that implies irrecoverable loss of phase. It is also interesting to include inhomogeneous broadening in hybrid systems explicitly in order to investigate possible phase revivals, which can be harvested by photon echo spectroscopy methods.[314] A population of molecules can exist that is described by a distribution of energies having a central transition energy. Each molecule is detuned from this central energy by some amount due to inhomogeneous broadening, which can result from conditions such as Doppler shift in individual gas molecules or variations in electric field from point to point in solids. Given such distribution of molecular transition energies, each individual molecule oscillates at a frequency that is slightly different from the others after the system is pumped by a strong incident pulse. As a result, all of the molecules begin to oscillate in phase at first, but eventually they dephase within a characteristic inhomogeneous lifetime and, if left alone, never re-phase again resulting in free induction decay.[282] However, for times less than the natural lifetime of the molecule, each molecule is still oscillating. These oscillations are lost when including pure dephasing as a simple damping. One can in principle invert the dephasing process by applying a second pump in the form of a π-pulse. The oscillations then all run in reverse, resulting in a subsequent rephasing. The ensemble of molecules exhibits a non-zero macroscopic polarization once again eventually emitting radiation known as a photon echo signal as schematically illustrated in Fig. 18a. This technique is widely used in chemistry and is referred to as photon echo spectroscopy.[315] Inhomogeneous effects due to variations in molecules' surroundings cause each molecule to oscillate at a slightly different frequency than the others, and photon echo spectroscopy removes this effect. Any remaining dephasing cannot be reveresed by the echo technique, and is revealed as diminished intensity of the echo.[316] For example, as the delay in applying the π-pulse increases, the natural lifetime of the emitters causes all of their oscillations to decrease, resulting in an echo with lower intensity. Thus a time and frequency structure of a detected photon echo contains important information about the probed system. Additionally, the recovery of a time signal after dephasing offers prospects for memory storage. In Ref. [317], the optical properties are copied to a spin system whose lifetime is much longer than that of the optical system thus extending the duration of the system's memory.

Recently[318] the two-pulse photon echo spectroscopy was applied to hybrid systems. The semiclassical theory is expanded to include inhomogeneous broadening of molecules explicitly. When molecules placed in a close proximity to plasmonic structures (such as periodic arrays of slits or nanoparticles) and are resonant to a plasmonic mode a double-peaked time signal of photon echo is observed. The corresponding Husimi transformation (time-frequency map of a time signal[319]) is shown in Fig. 18b. One can see that the time-frequency map exhibits two frequencies between 140 – 190 fs. Upon careful examination it was found that such two frequencies presented in the photon echo signals correspond to hybrid states of the system, *i.e.* upper and lower polaritons. It is interesting to note that in all numerical experiments performed in Ref. [318] the frequency of the upper polariton was always dominant in echo spectra as seen in the Husimi transformation of the time signal in Fig. 18b. It was also shown that the unique signature of upper and lower polaritons in photon echoes was highly sensitive to exciton-plasmon coupling making it a great tool for ultrafast optical probes of nanomaterials to further scrutinize fundamentals of light-matter interaction at the nanoscale.

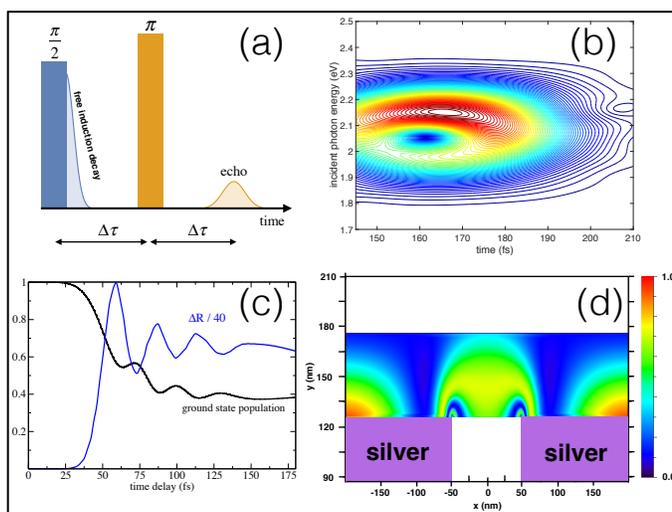

**Figure 18**. Panel (a) schematically shows the dynamics of the two-pulse photon echo spectroscopy. Panel (b) shows the Husimi transformation of the time signal of photon echo detected on the output side of the periodic slit array covered by a 20 nm thin resonant molecular layer. Panel (c) shows the ensemble average ground state population of molecules (black line) and the change in reflection as functions of the time delay between the pump and the probe pulses. Panel (d) shows the instantaneous spatial distribution of the ground state population of molecules inside molecular layer placed on top of the slit array during strong laser pulse excitation.

The resonant pump-probe nonlinear spectroscopy applied to hybrid nanomaterials can reveal interesting features of the electromagnetic energy exchange between plasmons and excitons.[199, 283] We apply the numerical procedure outlined in Section 3 to obtain time resolved spectra modified by a strong optical pump driving the hybrid system comprised of the periodic array of slits in a silver film covered by a 10 nm thin layer of resonant molecules. The pump pulse is 180 fs long and with the amplitude high enough to drive a single molecule through several Rabi cycles. The reflection is calculated at the upper polariton frequency as a function of the pump-probe delay. The change of the reflection with respect to its unperturbed value is shown in Fig. 18c along with the ensemble average population of the molecular ground state. There is a clear correspondence between the oscillations of the ground state population and the changes in reflection induced by the pump. (A similar correlation was found for changes in transmission.) The observed oscillations in the transient spectra are due to quantum transitions between ground and excited states in individual molecules. It is also important to examine the role of dephasing in hybrid materials. This property is greatly affected by the inhomogeneous electric field due to SPPs, as is evident from Fig. 18c. One of the causes of the fast decay of Rabi oscillations (in addition to pure decoherence and relaxation caused by interaction with the metallic system, which is explicitly included in the model) is spatially dependent excitation of the molecular layer, resulting from the strong gradient of the electric field induced mainly by surface plasmons. Different parts of the molecular ensemble therefore experience different local fields. This effect in turn changes the plasmon dynamics, influencing not only the amplitude of Rabi oscillations but also the Rabi period. It should be noted that the molecular layer influences the near-field via the polarization current, which results in even faster dephasing. Simulations with a thicker molecular layer confirm the aforementioned conclusion, i.e. the Rabi oscillations decay significantly faster when a 10 nm thick molecular layer is replaced by a layer with a thickness of 50 nm.[199]

It is informative to examine how molecules undergo Rabi cycling under the influence of a strong pump at different spatial locations near metal interface. As pointed out above, strong electromagnetic field gradients result in a highly inhomogeneous excitation of molecules. This is demonstrated in Fig. 18d where we plot a snap-shot of the spatial map of molecules in the ground state taken at the time corresponding to the maximum of the pump. A clear strong spatial variation of the ground-state population is seen. In fact, the variations of the ground state population are oscillations whose wavelength varies somewhat over the region of molecules. One might surmise that the wavelength of these oscillations is on the order of that of the pump, but it is actually significantly smaller. The retardation effects are revealed to be the cause of the spatial modulations of the ground-state population: the temporal oscillations of the ground state probability are similar between adjacent spatial points, but slightly shifted.[320] Hence at a given time, adjacent points have slightly different values of ground state probability. For larger pump amplitude, each molecule will undergo Rabi flopping more rapidly in time. Thus a given phase shift between adjacent locations will lead to a larger shift in the ground state population between those two locations. For a slower group velocity, the pump takes longer to reach an adjacent point, causing the temporal oscillations between two adjacent points to acquire a larger phase difference. The very reason for such spatial variations in the ground state population is the strong coupling between molecules and the corresponding plasmon mode resulting in a small group velocity.

## 5. Summary and outlook

Exciton-plasmon interaction and the laboratory observation of its consequences have attracted much attention in recent years. In this account we have briefly reviewed the main observations in which this coupling is expressed, described models for its description and presented the state of the art methodology for numerical simulations of its manifestations in key experimental observations. At the same time, we have emphasized the common origin of many of the observed phenomena and similar observations in other settings. This common origin is exemplified by the typical characteristics of "strong coupling" phenomena, whose manifestations as an avoided crossing between free exciton and free plasmon dispersion lines is qualitatively similar to many other phenomena including plasmon hybridization. Another example is provided by the similarity between exciton-plasmon coupling and coupling of molecular excitations to electromagnetic cavity modes. While these similarities are evident, it is important to keep in mind important practical differences. Even though avoided crossing is pervasive in many quantum and classical systems, its observation in exciton-plasmon systems is facilitated by the fact that many experimental setups make it possible to control the relevant excitation energy by exploring the frequency-wavevector dispersion properties. Furthermore, the appearance of strong coupling effects between collective modes of a many-body system is not to be taken for granted and limits to their observation must depend on coherence destroying processes. The study of this balance and its manifestations in different systems is still ahead of us. Also ahead of us, both on the experimental and theoretical-computational fronts is the exploration of the vast modes of behaviors that may be associated with the nonlinear response of such systems.

The systems involved in these studies are quite complex, and while simple models have been useful for qualitative understanding, their detailed investigations require numerical simulations. We have described the state of the art of these simulations and provided sample codes for interested readers to explore and utilize. In their simplest form, simulations consider the optical response of inhomogeneous dielectric systems, with the most common tool being numerical solvers based on FDTD method to integrate the corresponding Maxwell equations. We have described in detail a more advanced approach based on coupled Liouville-Maxwell equations, where the radiation field and its propagation in the metallic domains of the system are described as before by the Maxwell equations applied to the given inhomogeneous medium in which the molecular oscillating dipoles appear as source currents, while the molecular response is described by Liouville-Bloch equations that account for the quantum evolution of the molecular sub-system under the effects of the local electromagnetic field. This level of description can account for dephasing effects in and nonlinear response of the molecular subsystem, while still provide the versatility needed to treat realistic systems of different sizes and shapes.

Looking back, even in just the last decade the field has made an enormous progress in both experimental and numerical methodologies as well as in the discovery of new modes of behaviors of these interesting systems. Much can be achieved by further studying such systems on this level of sophistication, in particular in exploring their nonlinear response behavior and transient response phenomena. With respect to the latter we note that while avoided crossing of exciton and plasmon dispersion peaks is now often observed in such systems, its temporal counterpart, the observation of Rabi oscillations, has been demonstrating only once. Clearly, a better time resolution and/or better control on the preparation of the initially excited state is needed for such observations, and numerical simulations can play an important role in establishing the needed conditions.

Focusing on the numerical methodology, several directions for needed developments are evident. First and foremost is the classical description of the radiation field. While much can be achieved in a classical description it is obvious that purely classical description cannot account for the often observed system response at radiation wavelengths different from the incident field.[xi] Obviously, a full quantum description of the radiation field is not practical (except in cavities that support one or just a few modes), however some of the relevant quantum consequences can perhaps be accounted for (see Eq. (19) and the associated discussion).

Another issue that requires further study is the use of the mean field approximation (see Eqs. (30) - (32)) for the molecular response in our numerical modeling. Although numerical results based on this approximation do show coherent response, it is not clear at present how much is lost with respect to the description of coherent response by using this approximation. For example, it is known that excitonic energy transport can change its nature from coherent to diffusive due to thermal and disorder effects. It is

---

[xi] This statement holds in the linear response regime, but this cast doubt also on the integrity of nonlinear response signal obtained in a classical description.

certainly of interest to study this transition in the presence of coupling to a plasmonic interface, however it is not evident whether such a study can be done on the mean field level.

Finally, the optical response of metal-molecule interfaces often involves, in addition to collective molecular and metal excitations, also electron transfer between these subsystems. Indeed, the consequence of this type of charge transfer phenomena following optical excitation of such interfaces has become a subject of intense recent studies. The numerical description of such phenomena is obviously beyond the capabilities of the presently available numerical methodologies. It will be interesting to explore the possibility to apply methodologies based on the Semiconductor Bloch equations[321, 322] for such problems.

The study of coupled exciton-plasmon systems has come a long way in recent years. With a host of open issues, interesting observations and the promise of technological applications we expect it to remain active and fruitful for a substantial time.

## 6. Acknowledgements

The authors are thankful to Prof. Adi Salomon for fruitful discussions. M.S. is grateful to the financial support by the Air Force Office of Scientific Research under Grant No. FA9550-15-1-0189. Both M.S. and A.N. would like to thank Binational Science Foundation for generous financial support through collaborative Grant No. 2014113. A.N also acknowledges financial support from the University of Pennsylvania.